\documentclass[journal]{IEEEtran}
\usepackage{lineno}
\usepackage{cite}
\usepackage{hyperref}
\usepackage{amsmath,amssymb,amsfonts}
\usepackage{amsmath}
\usepackage{amsthm}

\usepackage{array}
\usepackage{graphicx}
\usepackage{textcomp}
\usepackage{xcolor}
\usepackage{graphicx}
\usepackage{float}
\usepackage{subfigure}
\usepackage{amsmath}
\usepackage{amsfonts,amssymb}
\usepackage{mathrsfs}
\usepackage{mathtools}
\usepackage{multirow}
\usepackage{algpseudocode}
\makeatletter
\newif\if@restonecol
\makeatother

\usepackage[linesnumbered,ruled,vlined]{algorithm2e}
\usepackage{algpseudocode}

\usepackage{setspace}
\usepackage{footmisc}
\usepackage[justification=centering]{caption}

\usepackage{mathtools}
\usepackage{dsfont}
\usepackage{bbm}

\newtheorem{corollary}{Corollary}

\theoremstyle{definition}
\newtheorem{theorem}{Theorem}

\newtheorem{lemma}{Lemma}

\makeatletter
\newcommand{\biggg}{\bBigg@{3}}
\newcommand{\Biggg}{\bBigg@{3.5}}
\makeatother
\usepackage{bm}
\usepackage{comment}

\newcommand{\angel}[1]{\noindent { {{$\blacktriangleright$
   {\textsf{[Angel]: {\color{red}#1}}} $\blacktriangleleft$}}}}
\newcommand{\hao}[1]{\noindent { {{$\blacktriangleright$
   {\textsf{[Hao]: {\color{blue}#1}}} $\blacktriangleleft$}}}}
\makeatother
\begin{document}

\title{Two-Timescale Design for Downlink Multiuser Transmission with
Dynamic Metasurface Antennas}

\author{Hao Xu, \emph{Graduate Student Member, IEEE},  Angel Lozano, \emph{Fellow, IEEE}, and Hongwen Yang, \emph{Member, IEEE}

\thanks{The work of Hao Xu and Hongwen Yang was supported in part by the BUPT Excellent Ph.D. Students Foundation under Grant CX2023149, in part by the China Scholarship Council, and in part by 5G Evolution Wireless Air interface Intelligent R\&D and Verification Public Platform Project (Grant. 2022-229-220).
The work of Angel Lozano was supported by AGAUR and by the Maria de Maeztu Units of Excellence Programme CEX2021-001195-M funded by MICIU/AEI/10.13039/501100011033.
}
\thanks{Hao Xu and Hongwen Yang are with the School of Information and Communication Engineering, Beijing University of Posts and Telecommunications, Beijing 100876, China (e-mail: Xu\_Hao@bupt.edu.cn; yanghong@bupt.edu.cn). Angel Lozano is with the Department of Engineering, Universitat Pompeu Fabra, Barcelona 08018, Spain (e-mail: angel.lozano@upf.edu).}
}

\maketitle

\begin{abstract}
Dynamic metasurface antennas (DMAs) promise to relieve massive multiple-input multiple-output architectures from their high energy consumption and hardware costs.
This paper proposes a two-timescale design for downlink multiuser transmission via DMAs, a design that balances pilot overhead, complexity, and spectral efficiency.
At the onset of each frame, the DMA coefficients are configured based only on statistical channel-state information (CSI), a process for which the paper introduces an optimization framework that is shown to outperform the widely used stochastic successive convex approximation method. Then, within each frame, the digital precoder is updated at each slot, based on the optimized DMA coefficients and the effective lower-dimensional instantaneous CSI. The weighted minimum mean-squared error method is applied for this short-term optimization and, for the special case of single-user transmission, a closed-form solution for the digital precoder is provided.
Performance evaluations demonstrate that the proposed two-timescale design can be an attractive ingredient for future wireless networks.
\end{abstract}

\begin{IEEEkeywords}
Dynamic metasurface antennas, multiuser communication, multiple-input single-output, two-timescale design.
\end{IEEEkeywords}

\begin{figure*}[!h]
\centering
\includegraphics[width=0.75\textwidth]{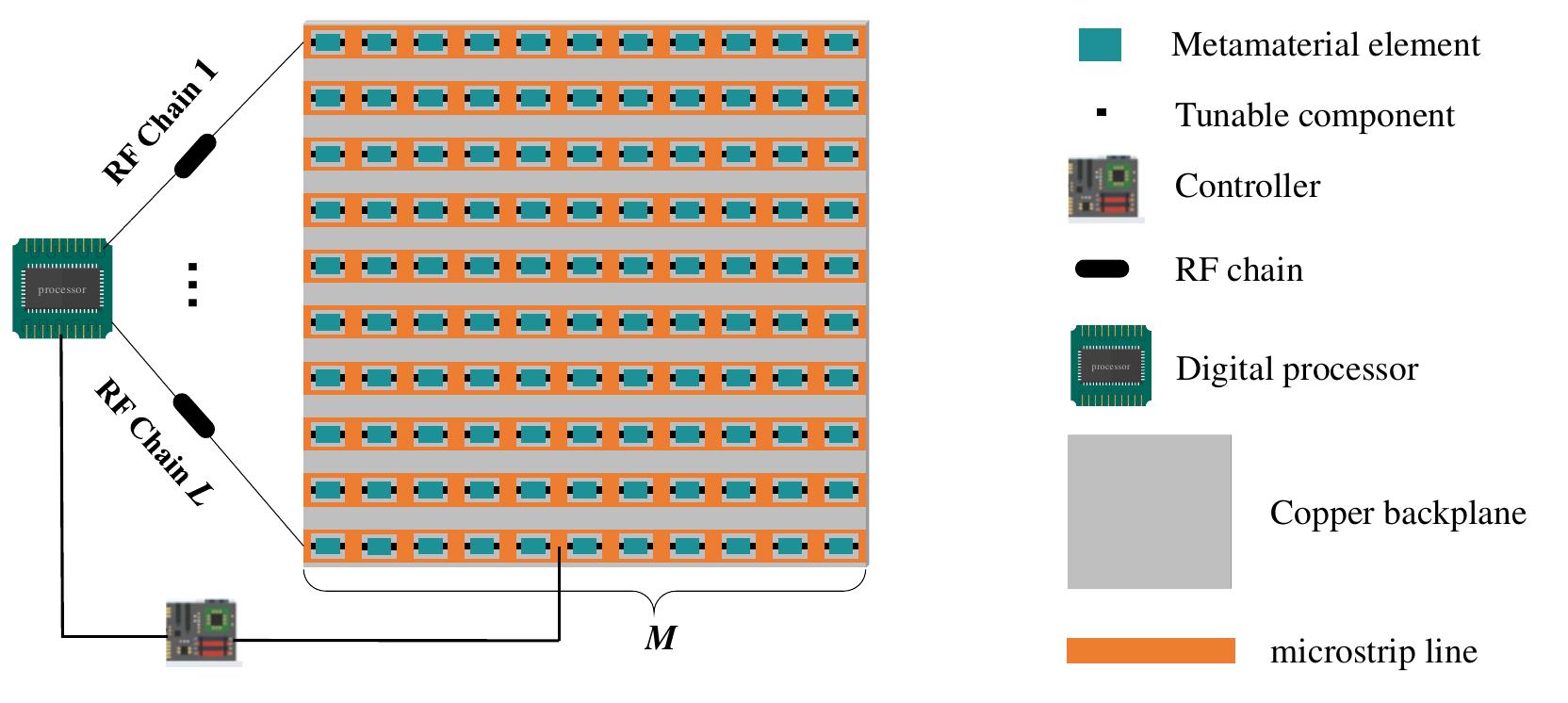}
\caption{DMA architecture.}
\label{system_model}
\end{figure*}

\section{Introduction}

Massive multiple-input multiple-output (MIMO) has been shown to improve spectral efficiency and throughput \cite{Larsson2014,Andrews2014,Saad2020,Tataria2021}. However, its implementation faces challenges, including high energy consumption, hardware costs, and deployment constraints, which become even more critical in the context of the extremely large-scale MIMO (XL-MIMO) envisioned for future 6G networks \cite{Rial2016,Mo2017}. To address these issues, recent research has focused on cost-effective and scalable embodiments. Among these, dynamic metasurface antennas (DMAs), empowered by advances in metamaterials, have emerged as a promising candidate \cite{Shlezinger2021,Jabbar2024}.

A DMA is a planar array of radiating metamaterial elements that are integrated onto waveguides connected to radio-frequency (RF) chains \cite{Shlezinger2019}. Each element operates as a resonator whose frequency response is characterized by the Lorentz model \cite{Smith2017}. By independently tuning the electromagnetic properties of individual elements, programmable control over both transmitted and received signals can be achieved, hence the descriptor ``dynamic''. DMA-based base stations (BSs) inherently realize hybrid analog-digital beamforming. In contrast to traditional hybrid architectures, though, DMA-based BSs eliminate the need for analog combining circuits such as phase shifters (PSs). Rather, the tuning of DMA elements entails simpler components, such as varactor diodes. As a result, DMAs offer lower hardware cost and power consumption than PS-based arrays \cite{Shlezinger2021}.

Owing to the aforementioned advantages, DMA-assisted systems have received considerable research attention. In \cite{Shlezinger2019}, the achievable sum-rate of a DMA-assisted uplink multiuser MIMO setup was analyzed, and this work was later extended to MIMO-OFDM in \cite{Wang2021}. A corresponding downlink analysis can be found in \cite{Wang2019}. Then, \cite{Kimaryo2023} proposed an alternating algorithm to jointly optimize the transmit precoders and DMA weights on the basis of the weighted sum-rate. The energy efficiency was studied for a DMA-assisted downlink multiuser setup in \cite{Chen2025}, and for a DMA-assisted RF wireless power transfer system in \cite{Azarbahram2024}. Additionally, \cite{Zhang2021} investigated the achievable rate of a near-field downlink multiuser multiple-input single-output system employing DMAs. Building upon this work, \cite{Zhang2022} conducted a comprehensive comparison between DMA and PS-based hybrid beamforming architectures. Focusing on near-field conditions, \cite{Xu2024} examined the uplink sum rate in wideband systems and \cite{Zhang2025} investigated the beam focusing for wireless power transfer systems. In turn, \cite{Rezvani2024} studied the channel estimation problem for DMAs.

All the aforementioned studies assume perfect instantaneous channel state information (iCSI) at the transmitter. However, due to the large-scale structure of DMAs (i) acquiring such iCSI would entail hefty channel estimation overheads, and (ii) real-time optimizations based on iCSI would incur high complexity, being outright unfeasible in high-mobility scenarios. To address these challenges, a number of studies have proposed designs based solely on statistical CSI (sCSI). For instance, \cite{Xu2025} aimed at maximizing the ergodic spectral efficiency of both uplink and downlink using only sCSI.
In \cite{Zhang2024}, a hybrid uplink multiuser system assisted by both a DMA and a reconfigurable intelligent surface (RIS) was investigated under sCSI. The corresponding energy efficiency was evaluated in \cite{You2023}. What emerges from these works is that designs based only on sCSI inevitably lead to performance degradation relative to iCSI-aware methods, especially in channels exhibiting strong multipath propagation \cite{Xu2025}.

With a view to achieving the complexity reductions of sCSI while preserving much of the performance of iCSI, this paper investigates a two-timescale design. In it, the DMA coefficients are optimized only at the beginning of each frame, based on sCSI. Given these optimized DMA coefficients and a low-dimensional iCSI, the digital precoder is then updated at the beginning of each slot within the frame. This two-timescale framework has been adopted for RIS-aided \cite{Zhao2021, Xu2023} and PS-based hybrid beamforming architectures \cite{Liu2018, Cai2020, Liu2025}, but, to the best of our knowledge, had not yet been studied for DMAs.
To assess its potential in this context, this paper considers a rather general correlated Rayleigh fading channel, under which the performance degradation resulting from the use of sCSI is pronounced \cite{Xu2025}. The contributions of the paper are as follows:
\begin{itemize}
  \item A two-timescale design is proposed for DMA-based downlink transmission to balance signaling overhead, complexity, and spectral efficiency.
  \item 
An optimization framework is proposed to obtain the long-term DMA coefficients based on sCSI, one that outperforms the widely adopted stochastic successive convex approximation (SSCA). The short-term digital precoder, in turn, is derived from those DMA coefficients and the effective iCSI via the weighted minimum mean-squared error (WMMSE) method.
\item For the important special case of single-user transmission, a closed-form solution is provided for the short-term digital precoder.
\item Performance evaluations are set forth, demonstrating that the proposed method outperforms existing schemes and highlighting the practicality of the two-timescale design.
\end{itemize}

The paper is organized as follows. Sec. \uppercase\expandafter{\romannumeral2} reviews the DMA architecture, introduces the system model, and formulates the two-timescale design. Sec. \uppercase\expandafter{\romannumeral3} presents the optimization framework for long-term DMA coefficients and the WMMSE-based method for short-term digital precoding. Sec. \uppercase\expandafter{\romannumeral4} specializes these optimizations to the single-user scenario, Sec. \uppercase\expandafter{\romannumeral5} assesses the performance, and Sec. \uppercase\expandafter{\romannumeral6} concludes the paper.

\textbf{Notation:} Boldface lower-case letters ($\boldsymbol{x}$) and upper-case letters ($\boldsymbol{X}$) denote column vectors and matrices, respectively. For a matrix $\boldsymbol{X}$, $[\boldsymbol{X}]_{i,j}$ denotes the element in the $i$th row and $j$th column, $\|\boldsymbol{X}\|_2$ denotes the spectral norm, and $\|\boldsymbol{X}\|_\text{F}$ denotes the Frobenius norm. For a vector $\boldsymbol{x}$, $[\boldsymbol{x}]_i$ denotes the $i$th element and $\|\boldsymbol{x}\|$ denotes the Euclidean norm.
Moreover, $(\cdot)^{\mathsf T}$ and $(\cdot)^{*}$ respectively denote the transpose and transpose conjugation operation,
$\boldsymbol{0}_N$ denotes the $N \times 1$ zeros vector, $\angle(\cdot)$ denotes the phase extraction operator, $\mathsf{diag}(\boldsymbol{X})$ and $\mathsf{diag}(\boldsymbol{x})$ respectively denote the diagonal extraction operator and the expansion from vector to diagonal matrix, while $\odot$ denotes the Hadamard product, $\otimes$ denotes the Kronecker product, and $\Re(\cdot)$ indicates the real part.



\section{System Model and Problem Formulation}

\subsection{DMA Architecture}

As shown in {\figurename} \ref{system_model}, the DMA considered in this paper is a planar array of radiating metamaterial elements arranged along microstrip lines. Each such microstrip is connected to an individual RF chain. The DMA contains $L$ RF chains and microstrips, each feeding $M$ elements. The total number of DMA elements is $N=LM$. Each element, featuring tunable components such as varactors, can be modeled as a resonant circuit. By adjusting the bias voltage of the varactors through an integrated controller, the electromagnetic response of the elements can be dynamically controlled.


As the signal propagates along the microstrips, its phase varies according to the wavelength and the position of each element. Additionally, there is attenuation. Altogether, the propagation to the $m$th element ($m=1,\dots,M$) on the $\ell$th ($\ell=1,\dots,L$) microstrip is characterized by
\begin{align}\label{h}
a_{\ell, m}=e^{-d_{\ell, m}\left(\alpha+\jmath \frac{2\pi\sqrt{\varepsilon}}{\lambda}\right)},
\end{align}
where $\alpha$ is the attenuation coefficient, $\varepsilon$ is the dielectric constant of the microstrip lines, $\lambda$ is the free-space wavelength, and $d_{\ell,m}$ is the distance from the $m$th element to the $\ell$th RF port.


As the resonance bandwidth is broader than any signal bandwidth foreseen for 6G, the response of each element can be regarded as frequency-flat \cite{Shlezinger2021},
taking the form \cite{Smith2017}
\begin{align} \label{cons_Q}
q_{\ell,m} =\frac{\jmath+ {\rm{e}}^{\jmath\theta_{\ell,m}}}{2}
\end{align}
with the configuration controlled via the phase shifts $\{ \theta_{\ell,m} \}$.

Let $\boldsymbol{s}=[s_1,\dots,s_L]^{\mathsf T} \in \mathbb{C}^{L \times 1}$ where $s_\ell$ is the signal produced by the $\ell$th RF port. Then, at the $m$th element on the $\ell$th microstrip,
\begin{align}\label{x_lm}
z_{\ell,m}=q_{\ell,m} a_{\ell,m} s_\ell . 
\end{align}
This relationship can be vectorized into
\begin{align} \label{DMA}
\boldsymbol{z}=\boldsymbol{QAs},
\end{align}
where
\begin{align}
\boldsymbol{z} & = [z_{1,1},\dots,z_{1,M},\dots,z_{\ell,m},\dots,z_{L,M}]^{\mathsf T} \in\mathbb{C}^{N\times 1} \\
\boldsymbol{q} & =[q_{1,1},\dots,q_{\ell,m},\dots,q_{L,M}]^{\mathsf T}\in\mathbb{C}^{N\times 1}
\end{align}
and $\boldsymbol{Q} =\mathsf{diag}(\boldsymbol{q})\in \mathbb{C}^{N\times N}$,
while
$\boldsymbol{A}\in \mathbb{C}^{N \times L}$ with
\begin{align}
[\boldsymbol{A}]_{i,j}=\left\{\begin{array}{ll}
a_{j, \text{mod}(i-1,M)+1} & (j-1)M+1\leq i \leq jM \\
0 & \text{otherwise}
\end{array} . \right.
\end{align}

\subsection{System Model}

With $K$ single-antenna users, the observation at the $k$th user is given by
\begin{align}\label{MU}
y_k=\boldsymbol{g}_k^{*}\boldsymbol{QAWx}+n_k,
\end{align}
where $\boldsymbol{g}_k\in \mathbb{C}^{N \times 1}$ is the channel from the BS, $\boldsymbol{x}$ is the transmit vector,
and
\begin{align}
\boldsymbol{W}=[\boldsymbol{w}_1 \cdots \boldsymbol{w}_K] \in {\mathbb{C}}^{L\times K}
\end{align}
is the digital precoding matrix given ${\boldsymbol w}_k\in {\mathbb{C}}^{L\times 1}$ as the precoding vector for user $k$. In turn, $n_k \sim \mathcal{N}_\mathbb{C}(0,\sigma_k^2)$ is white Gaussian noise of power $\sigma_k^2$.

A correlated Rayleigh fading channel model is invoked, whereby
\begin{align}
\boldsymbol{g}_k=\boldsymbol{R}_k^{1/2}\tilde{\boldsymbol{g}}_k,
\end{align}
where $\boldsymbol{R}_k$ is the spatial correlation matrix for user $k$ while $\tilde{\boldsymbol{g}}_k\sim\mathcal{N}_\mathbb{C}(0,\boldsymbol{I}_N)$. The $K$ channels are mutually independent and, for notational compactness, they are assembled into
\begin{align}
\boldsymbol{G} = [\boldsymbol{g}_1  \cdots  \boldsymbol{g}_K] \in \mathbb{C}^{N \times K} .
\end{align}

\subsection{Transmission Protocol}

\begin{figure}[!t]
\centering
\includegraphics[height=0.18\textwidth]{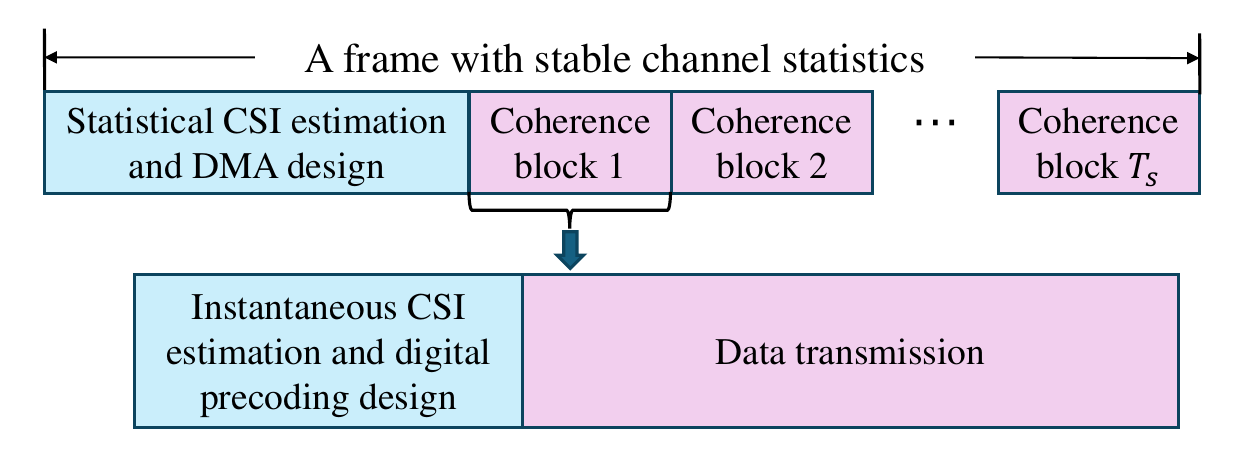}
\caption{Two-timescale transmission protocol.}
\label{transmission_protocol}
\end{figure}

At the beginning of each time frame, the statistics of the high-dimensional channel $\boldsymbol{G}$ are estimated, which amounts to gathering the $K$ correlation matrices \cite{Mestre2008}. These, along with $\boldsymbol{A}$ and $\boldsymbol{Q}$, are used to configure the DMA coefficients, which are upheld over the entire frame (see Fig. \ref{transmission_protocol}). Then, the digital precoder for every user $k$ is designed at every slot within the frame based on those DMA coefficients and the lower-dimensional effective channel
\begin{align}
\boldsymbol{h}_k=\boldsymbol{A}^{*}\boldsymbol{Q}^{*} \boldsymbol{g}_k \in \mathbb{C}^{L\times 1}.
\end{align}
The $K$ effective channels can be learned by the BS through standard procedures: by observing uplink pilots if the system is time-division duplexed (TDD), or by having the users report back from their observations of unprecoded downlink pilots if the system is frequency-division duplexed (FDD) \cite{Sun2002}.
Importantly, these channels are of dimension $L \ll N$. The number of uplink pilots is proportional to $K$ in the TDD case whereas the number of unprecoded downlink pilots is proportional to $K L$ in the FDD case.

Every user $k$, for its part, can estimate its own end-to-end channel $\boldsymbol{g}^*_k \boldsymbol{QA}\boldsymbol{w}_k$ by observing downlink precoded pilots. These $K$ channels are scalar-valued, hence the number of precoded downlink pilots is proportional to $K$.

\subsection{Problem Formulation}

With user $k$ knowing $\boldsymbol{g}^*\boldsymbol{QA}\boldsymbol{w}_k$, the maximization of the ergodic sum spectral efficiency can be formulated as \begin{subequations} \label{P_MU}
\begin{align}
& \max_{\boldsymbol{Q}}~\mathbb{E}_{\{\boldsymbol{g}_k\}}  \! \left[\max_{\{\boldsymbol{w}_k\}} ~\sum_{k=1}^{K}\log_2(1+\gamma_k) \right]\label{obj_MU} \\
& \; {\rm{s.t.}}  \;\, \eqref{cons_Q} , \, \sum_{k=1}^{K}\|\boldsymbol{QA}\boldsymbol{w}_k\|^2 \leq P_\text{t}, \label{power_constraint_MU}
\end{align}
\end{subequations}
where $P_\text{t}$ is the transmit power while
\begin{align}\label{SINR_k}
\gamma_k=\frac{|\boldsymbol{g}_k^{*} \boldsymbol{Q}\boldsymbol{A} \boldsymbol{w}_k|^2}{\sum_{i\neq k}|\boldsymbol{g}_k^{*} \boldsymbol{Q}\boldsymbol{A} \boldsymbol{w}_i|^2+\sigma_k^2}
\end{align}
is the signal-to-interference-plus-noise ratio (SINR) at user $k$.

The problem in \eqref{P_MU} exhibits a nested stochastic structure and nonconvex constraints. The outer expectation over channel realizations, coupled with the inner maximization over precoders, results in a two-layer stochastic optimization problem that is analytically intractable. Moreover, the joint power constraint couples the optimization variables, and the ergodic sum spectral efficiency is nonconcave in the optimization variables.
The problem is altogether challenging, and
an algorithmic framework to tackle it is unveiled next.

\section{Proposed Two-Timescale Design}



\subsection{Short-Term Design}  \label{III_A}

For given DMA weights $\boldsymbol{Q}$ and effective channels $\{ \boldsymbol{h}_k \}$, the optimization problem with respect to $\boldsymbol{W}$ is
\begin{subequations} \label{P_W}
\begin{align}
& \max_{\boldsymbol{W}} \sum_{k=1}^{K}\log_2  \! \left(1+\frac{|\boldsymbol{h}_k^{*}\boldsymbol{w}_k|^2} {\sum_{i\neq k}|\boldsymbol{h}_k^{*} \boldsymbol{w}_i|^2+\sigma_k^2}\right) \label{P_obj_wk} \\
& \, {\rm{s.t.}}  \;\, \sum_{k=1}^{K}\|\boldsymbol{QA}\boldsymbol{w}_k\|^2 \leq P_\text{t} , \label{power_constraint_w}
\end{align}
\end{subequations}
which retains the challenges of the nonconcave objective function and coupled constraint. Through the WMMSE framework, the above can be transformed into the equivalent problem\footnote{The logarithm base in the transformed problems does not affect the optimizer. Here, natural logarithms are used for notational simplicity.} \cite{Shi2011}
\begin{subequations}\label{WMMSE}
\begin{align}
& \min_{\boldsymbol{W},\{ u_k \},\{\xi_k\}} \,  \sum_{k=1}^{K}  \Big(\xi_k
\, \mathbb{E} \big[ (u_k^{*}y_k-x_k)(u_k^{*}y_k-x_k)^{*} \big] \label{obj_WMMSE} \\
& \qquad\qquad\qquad\qquad -\log\xi_k\Big) \notag \\
 & \;\, {\rm{s.t.}} \;\, \sum_{k=1}^{K}\|\boldsymbol{QA}\boldsymbol{w}_k\|^2 \leq P_\text{t} ,
\end{align}
\end{subequations}
whose constraint is convex and whose objective function, although not jointly convex in all the optimization variables, is separately convex in each of them.
Therefore, by alternately optimizing $\{u_k\}$,$\{\xi_k\}$, and $\boldsymbol{W}$ until convergence, a locally optimal solution can be obtained for $\boldsymbol{W}$.

Specifically, $\{ u_k \}$ are first optimized while keeping $\{\xi_k\}$, and $\boldsymbol{W}$ fixed. By setting the corresponding partial derivatives to zero,
one obtains
\begin{align} \label{opt_u}
u_k &=\frac{\boldsymbol{h}_k^{*}\boldsymbol{w}_k} {\sum_{i=1}^{K}|\boldsymbol{h}_k^{*} \boldsymbol{w}_i|^2+\sigma_k^2} .
\end{align}
Similarly, with
$\{u_k\}$ and $\boldsymbol{W}$ fixed, one obtains
\begin{align} \label{opt_xi}
\xi_k &=(1-u_k^{*}\boldsymbol{h}_k^{*}\boldsymbol{w}_k )^{-1}.
\end{align}
Finally, for given $\{ u_k \}$ and $\{\xi_k\}$, the optimization over $ \boldsymbol{w}_k $ takes the form
\begin{subequations}\label{WMMSE_W}
\begin{align}
& \min_{\boldsymbol{w}_k}  \boldsymbol{w}_k^{*}  \sum_{i=1}^{K} \xi_i  |u_i|^2 \boldsymbol{h}_i \boldsymbol{h}_i^{*}  \boldsymbol{w}_k-2\Re\big( \xi_k u_k \boldsymbol{w}_k^{*}\boldsymbol{h}_k\big) \label{obj_WMMSE_W} \\
& \; {\rm{s.t.}} \sum_{k=1}^{K}\|\boldsymbol{QA}\boldsymbol{w}_k\|^2 \leq P_\text{t}.
\end{align}
\end{subequations}
Letting $\boldsymbol{F}=\boldsymbol{QA}\in\mathbb{C}^{N\times L}$, this matrix is of full column rank and therefore
$\boldsymbol{F}^{*} \boldsymbol{F}$ is invertible. Lagrange multipliers can be applied to \eqref{WMMSE_W}, whose solution emerges (see Appendix \ref{Appendix_A}) as
\begin{align}\label{opt_W1}
\boldsymbol{w}_k = u_k \xi_k \left(\sum_{i=1}^{K} \xi_i |u_i|^2 \boldsymbol{h}_i \boldsymbol{h}_i^{*} + \lambda_{\text{w}} \boldsymbol{F}^{*} \boldsymbol{F} \right)^{\!\!-1} \!\! \boldsymbol{h}_k.
\end{align}
Here, $\lambda_{\text{w}}$ is the Lagrange multiplier associated with the power constraint, which can be obtained via the bisection method.



\subsection{Long-Term Design}

Armed with the sCSI embodied by the correlation matrices $\{\boldsymbol{R}_k\}$, the BS needs to solve \eqref{P_MU} in order to
optimize the DMA coefficients.
To tackle this challenging problem, existing research often resorts to the SSCA algorithm \cite{Zhao2021,Cai2020}, widely adopted for long-term analog beamformer design. This approach constructs an upper bound of the original objective function as a surrogate function  of $\boldsymbol{Q}$. By alternately generating instantaneous channel samples from the sCSI, updating the surrogate (including updating $\boldsymbol{W}$ through the WMMSE-based method), and optimizing $\boldsymbol{Q}$, the algorithm converges to a suboptimal solution. As it turns out, SSCA may suffer from
performance degradation due to the following:
\begin{itemize}
\item A limited number of channel samples might cause inaccuracies in the
stochastic approximation,
biasing the gradients and reducing the efficacy of the updates \cite{Zhao2021}.
\item
$\boldsymbol{Q}$ and $\boldsymbol{W}$ are not optimized under a unified power constraint, but rather in a decoupled fashion. Precisely, there is an outer optimization of $\boldsymbol{Q}$ subject only to \eqref{cons_Q}, but with no restriction in power. Then, the inner optimization of $\boldsymbol{W}$ is subject to a possibly stricter power constraint, as required to correct any excess in the outer optimization. 
\end{itemize}

To overcome these issues, a more robust long-term design is proposed that accommodates the joint power constraint. Specifically, $\boldsymbol{Q}$ and $\boldsymbol{W}$ are jointly optimized, yet only $\boldsymbol{Q}$ is retained; the digital precoders are discarded, and optimized at the short timescale as detailed in the previous section.
The joint optimization is
\begin{subequations} \label{P_sCSI}
\begin{align}
& \max_{\boldsymbol{Q},\boldsymbol{W}} \mathbb{E}_{\{ \boldsymbol{g}_k \}} \!  \! \left[ \sum_{k=1}^{K}\log_2 (1+\gamma_k ) \right]\label{obj_P_sCSI} \\
& \; {\rm{s.t.}} \;\, \eqref{cons_Q}, ~\sum_{k=1}^{K}\|\boldsymbol{QA}\boldsymbol{w}_k\|^2 \leq P_\text{t},\label{power_constraint_MU}
\end{align}
\end{subequations}
which is hampered by the lack of a closed form for the objective function.
To sidestep this obstacle, \eqref{obj_P_sCSI} can be approximated, as detailed in App. \ref{Appendix_B} and validated later, by
%
\begin{align} \label{MU_Approximation}
\sum_{k=1}^{K}\log_2\left(1+\bar{\gamma}_k\right),
\end{align}
where
\begin{align}
\bar{\gamma}_k=\frac{\boldsymbol{w}_k^{*} \boldsymbol{A}^{*}\boldsymbol{Q}^{*} \boldsymbol{R}_k \boldsymbol{QA}\boldsymbol{w}_k }{\sum_{i\neq k}^{K}\boldsymbol{w}_i^{*} \boldsymbol{A}^{*}\boldsymbol{Q}^{*} \boldsymbol{R}_k \boldsymbol{QA}\boldsymbol{w}_i+\sigma_k^2}.
\end{align}
It follows that problem \eqref{P_sCSI} is tightly approximated by
\begin{subequations} \label{P_appr}
\begin{align}
& \max_{\boldsymbol{Q},\boldsymbol{W}}  ~\sum_{k=1}^{K}\log_2 (1+\bar{\gamma}_k ) \label{obj_P_appr}\\
& \; {\rm{s.t.}} \;\, \eqref{cons_Q}, ~ \sum_{k=1}^{K}\|\boldsymbol{QA}\boldsymbol{w}_k\|^2 \leq P_\text{t},
\end{align}
\end{subequations}
where the remaining difficulties are the nonconvex objective function and the coupling of $\boldsymbol{Q}$ and $\boldsymbol{W}$ in the power constraint.
To deal with the first difficulty, \eqref{P_appr} can be transformed via fractional programming (see Appendix \ref{Appendix_C}) into the equivalent problem  
\begin{subequations}\label{P_FP}
\begin{align}
& \!\!\!  \max_{\boldsymbol{Q},\boldsymbol{W},\{\Gamma_k\},\{{\bm\beta}_k\}}  \sum_{k=1}^{K} \Big( \! \log(1+\Gamma_k)-\Gamma_k+(1+\Gamma_k) \mathcal{A}_k \Big) \label{P_FP_obj}\\
& \qquad\! {\rm{s.t.}}~ \eqref{cons_Q}, ~\sum_{k=1}^{K}\|\boldsymbol{QA}\boldsymbol{w}_k\|^2\leq P_\text{t}, \label{P_FP_cons}
\end{align}
\end{subequations}
where $\{\Gamma_k\}$ and $\{\bm\beta_k\}$ are auxiliary variables, and
\begin{align}
\mathcal{A}_k&=2 \, \Re \! \left(\bm\beta_k^{*} \boldsymbol{R}_k^{1/2} \boldsymbol{QA}\boldsymbol{w}_k \right)\nonumber\\
& \quad -\bm\beta_k^{*}\bm\beta_k \left( \sum_{i=1}^{K}\boldsymbol{w}_i^{*} \boldsymbol{A}^{*}\boldsymbol{Q}^{*} \boldsymbol{R}_k \boldsymbol{QA}\boldsymbol{w}_i +\sigma_k^2 \right).\nonumber
\end{align}
Notably, the objective function in \eqref{P_FP_obj} is separately concave over $\boldsymbol{Q},\boldsymbol{W},\{\Gamma_k\},\{{\bm\beta}_k\}$.

The second difficulty, i.e., the coupling of $\boldsymbol{Q}$ and $\boldsymbol{W}$ in the power constraint, motivates the use of an algorithm based on penalty dual decomposition (PDD), which is characterized by an embedded double-loop structure: the inner loop solves an augmented Lagrangian subproblem while the outer loop updates the penalty parameter or dual variable based on the constraint violation \cite{Shi2020}.


An additional auxiliary variable $\boldsymbol{P}=\boldsymbol{QAW}$ is introduced, which is subject to $\mathsf{tr}(\boldsymbol{P}\boldsymbol{P}^{*})\leq P_\text{t}$.
With that, \eqref{P_FP} becomes
\begin{subequations}\label{P_FP2}
\begin{align}
& \max_{
\stackrel{
\boldsymbol{Q},\boldsymbol{W},\boldsymbol{P}}
{\{\Gamma_k\},\{\bm\beta_k\}}
}
 \sum_{k=1}^{K} \Big( \! \log(1+\Gamma_k)-\Gamma_k+(1+\Gamma_k) \mathcal{B}_k \Big) \label{obj_P_FP2}\\
& \quad\, {\rm{s.t.}}~ \eqref{cons_Q},~\mathsf{tr}(\boldsymbol{P}\boldsymbol{P}^{*})\leq P_\text{t},~\boldsymbol{P}=\boldsymbol{QAW},
\end{align}
\end{subequations}
where
\begin{align}
\mathcal{B}_k=2 \, \Re \! \left(\bm\beta_k^{*} \boldsymbol{R}_k^{1/2} \boldsymbol{p}_k \right)-\bm\beta_k^{*}\bm\beta_k \left(\sum_{i=1}^{K}\boldsymbol{p}_i^{*} \boldsymbol{R}_k \boldsymbol{p}_i +\sigma_k^2\right) \nonumber
\end{align}
given $\boldsymbol{P}=[\boldsymbol{p}_1 \cdots \boldsymbol{p}_K]$. By moving the equality constraint as a penalty to the objective function, \eqref{P_FP2} is converted into the augmented Lagrangian form
\begin{subequations}\label{P_AL}
\begin{align}
& \min_{\boldsymbol{Q},\boldsymbol{W},\boldsymbol{P},\{\Gamma_k\},\{\bm\beta_k\}} -\sum_{k=1}^{K} \Big( \! \log(1+\Gamma_k)-\Gamma_k+(1+\Gamma_k) \mathcal{B}_k \Big)  \nonumber\\
& \qquad\qquad\qquad\quad +\frac{1}{2\rho}\left\|\boldsymbol{QAW}-\boldsymbol{P} +\rho{\bm\Psi}\right\| _\text{F}^2 \label{P_AL_obj}\\
& \quad\qquad\! {\rm{s.t.}}~ \eqref{cons_Q},~ \mathsf{tr}(\boldsymbol{P}^* \boldsymbol{P})\leq P_\text{t},
\end{align}
\end{subequations}
where $\bm\Psi=[\bm\psi_1  \cdots \bm\psi_K] \in \mathbb{C}^{N \times K}$ denotes the dual variable while $\rho >0$ is the penalty parameter. The coupling between $\boldsymbol{Q}$ and $\boldsymbol{W}$ is now relegated to the objective function.

In the PDD inner loop, the above augmented Lagrangian problem is solved by alternately optimizing over $\{\Gamma_k\}$, $\{\bm\beta_k\}$, $\boldsymbol{P}$, $\boldsymbol{W}$, and $\boldsymbol{Q}$, until convergence to a local optimum.
After initializing $\bm\Psi $ and $\rho$, the iterations proceed as follows.

\subsubsection{Step 1}
With all other variables fixed, $\{\Gamma_k\}$ and $\{\bm\beta_k\}$ can be optimized in closed form by setting partial derivatives to zero, giving
\begin{align}
\Gamma_k&=\frac{\boldsymbol{p}_k^{*} \boldsymbol{R}_k \boldsymbol{p}_k}{\sum_{i\neq k}^{K} \boldsymbol{p}_i^{*} \boldsymbol{R}_k \boldsymbol{p}_i+\sigma^2_k}, \label{opt_Gamma}\\
\bm\beta_k&= \left(\sum_{i=1}^{K} \boldsymbol{p}_i^{*} \boldsymbol{R}_k \boldsymbol{p}_i+\sigma^2_k\right)^{\!\!-1} \!\!\! \boldsymbol{R}_k^{1/2} \boldsymbol{p}_k. \label{opt_beta}
\end{align}

\subsubsection{Step 2}
Fixing all other variables and ignoring irrelevant terms, the subproblem of optimizing $\boldsymbol{p}_k$ can be written as
\begin{subequations}\label{P_P}
\begin{align}
& \min_{\boldsymbol{p}_k} ~ \boldsymbol{p}_k^{*} \bm\Omega \boldsymbol{p}_k-2\Re(\boldsymbol{p}_k^{*} \bm\phi_k ) \\
& \; {\rm{s.t.}}~ \sum\nolimits_{k=1}^{K}\|\boldsymbol{p}_k\|^2\leq P_\text{t},
\end{align}
\end{subequations}
given
\begin{align}
\bm\Omega=\sum_{i=1}^{K} (1+\Gamma_i)\bm\beta_i^{*} \bm\beta_i\boldsymbol{R}_i+\frac{1}{2\rho}\boldsymbol{I}_N
\end{align}
and
\begin{align}
\bm\phi_k=(1+\Gamma_k) \boldsymbol{R}_k^{1/2} \bm\beta_k + \frac{1}{2\rho}(\boldsymbol{QA}\boldsymbol{w}_k+\rho \bm\psi_k).
\end{align}
The above is a standard convex problem, whose optimal solution is
\begin{align} \label{opt_P}
\boldsymbol{p}_k =\left(\bm\Omega+\lambda_p \boldsymbol{I}_N\right)^{-1} \bm\phi_k
\end{align}
with $\lambda_\text{p} \geq 0$ being the Lagrangian multiplier, which can again be obtained through the bisection method.

\subsubsection{Step 3}
The subproblem of optimizing over $\boldsymbol{W}$ is 
\begin{align}
\min_{\boldsymbol{W}}~ \left\|\boldsymbol{QAW}-\boldsymbol{P}+\rho{\bm\Psi}\right\|_\text{F}^2,
\end{align}
with solution
\begin{align}\label{opt_W}
\boldsymbol{W} ={\boldsymbol{C}}^{-1} \boldsymbol{A}^{*} \boldsymbol{Q}^{*} (\boldsymbol{P}-\rho\bm\Psi),
\end{align}
where $\boldsymbol{C} = \boldsymbol{A}^{*} \boldsymbol{Q}^{*} \boldsymbol{QA}  \in \mathbb{C}^{L \times L}$ is a diagonal matrix whose $i$th diagonal entry equals
\begin{equation}
    [\boldsymbol{C}]_{i,i}= \!\!\!\!  \sum_{n=(i-1)M+1}^{iM} \!\! \left\vert[\boldsymbol{Q}]_{n,n}[\boldsymbol{A}]_{n,i}\right\vert^2.
\end{equation}

\subsubsection{Step 4}
The optimization of $\boldsymbol{Q}$ amounts to
\begin{align} \label{P_Q}
& \min_{\boldsymbol{Q}} \left\|\boldsymbol{QAW}-\boldsymbol{P}+\rho{\bm\Psi}\right\|_\text{F}^2 \\
& \; {\rm{s.t.}}~ \eqref{cons_Q},
\end{align}
which, disregarding irrelevant items, reduces to
\begin{subequations}\label{P_q}
\begin{align}
& \max_{\boldsymbol{q}} ~ \sum_{i=1}^{N} \Big(2 \, \Re\big( [\boldsymbol{q}]_i^{*} [\boldsymbol{a}]_i \big) -|[\boldsymbol{q}]_i|^2 [\boldsymbol{B}]_{i,i}\Big) \label{obj_P_q} \\
& \; {\rm{s.t.}}~ \eqref{cons_Q},
\end{align}
\end{subequations}
where $\boldsymbol{B}=\boldsymbol{A}\boldsymbol{W} \boldsymbol{W}^{*} \boldsymbol{A}^{*}$ and $\boldsymbol{a}=\mathsf{diag}\big((\boldsymbol{P}-\rho\bm\Psi) \boldsymbol{W}^{*}\boldsymbol{A}^{*}\big)$.
From \eqref{cons_Q},
\begin{align}
[\boldsymbol{q}]_i=\frac{\jmath+ {\rm{e}}^{\jmath\theta_i}}{2},
\end{align}
which, plugged into \eqref{obj_P_q}, yields
\begin{align}\label{opt_q}
\theta_i=\angle \Big([\boldsymbol{a}]_i-\jmath\frac{[\boldsymbol{B}]_{i,i}}{2}\Big),
\end{align}
where $\angle(\cdot)$ returns the angle.
The optimum $\boldsymbol{Q}$ then follows.


\vspace{5mm}

Let us now turn our attention to the outer loop, which begins by calculating the constraint violation as
\begin{align}\label{h}
h=\left\|\boldsymbol{QAW}-\boldsymbol{P}\right\|_\text{F}.
\end{align}
The value of $h$ determines whether to update $\bm\Psi$ or $\rho$. Precisely, when $h$ is below a threshold (which itself shrinks as the iterations progress), $\bm\Psi$ is updated; otherwise, $\rho$ is decreased. With that, the PDD method adaptively switches between the augmented Lagrangian and the penalty methods, gradually finding a  $\rho$ that ensures convergence.
The dual variable is updated via
\begin{align}
\bm\Psi^{t+1}=\frac{1}{\rho}(\boldsymbol{QAW}-\boldsymbol{P}) +\bm\Psi^{t}
\end{align}
and the penalty parameter as \cite{Shi2020}
\begin{align}
\rho^{t+1}=c_1\rho^t ,
\end{align}
where $t$ is the outer iteration number and $c_1$ is a scaling factor. The outer loop terminates once the constraint violation $h$ drops below a threshold.

The complete procedure is summarized in Algorithm \ref{Algorithm2}.
Recall that, as advanced, only the ensuing $\boldsymbol{Q}$ is retained, while
$\boldsymbol{W}$ is discarded as the precoders are optimized at the shorter timescale as detailed in Sec. \ref{III_A}.

\begin{algorithm}[!t]
    \caption{PDD-based method for solving \eqref{P_appr}}
    \label{Algorithm2}
    \textbf{Input}: $\{\boldsymbol{R}_k\}$;\\
    Initialize $\{\Gamma_k\},\{\bm\beta_k\},\boldsymbol{P}, \boldsymbol{W}, \boldsymbol{Q}$, dual variable $\bm\Psi$, constraint violation $h$, threshold $\epsilon$, $\eta$, scaling factors $c_1<1$, $c_2<1$, outer iteration index $t=0$, and penalty factor $\rho>0$\;
    \Repeat{$h< \epsilon$}
    {
    \Repeat{the objective function in \eqref{P_AL_obj} converges}
    {
    Update $\{\Gamma_k\},\{\bm\beta_k\},\boldsymbol{P}, \boldsymbol{W}, \boldsymbol{Q}$ based respectively on \eqref{opt_Gamma}, \eqref{opt_beta}, \eqref{opt_P}, \eqref{opt_W}, \eqref{opt_q};
    }
    Calculate $h$ in \eqref{h}\;
    \eIf{$h < \eta^t$}{
    Update $\bm\Psi^{t+1}=\frac{1}{\rho}(\boldsymbol{QAW}-\boldsymbol{P}) +\bm\Psi^{t}$\;} {
    Update $\rho^{t+1}=c_1\rho^t$\;
    }
    Update $\eta^{t+1}=c_2h$, $t=t+1$\;
    }
    \textbf{Output}: $\boldsymbol{Q}$.
\end{algorithm}

\subsection{Complexity}


For each inner loop iteration of Algorithm \ref{Algorithm2}, the number of multiply-and-accumulate operations to
optimize $\{\Gamma_k\},\{\bm\beta_k\},\boldsymbol{P}, \boldsymbol{W}$, $ \boldsymbol{Q}$ are respectively $\mathcal{O}(K^2N^2)$, $\mathcal{O}(K^2N^2)$, $\mathcal{O}(KN^3)$, $\mathcal{O}(NK)$, and $\mathcal{O}(NLK+N^2K)$. As for the outer loop, updating $\bm\Psi$ has complexity $\mathcal{O}(NK)$ while updating $\rho$ has complexity $\mathcal{O}(1)$. Altogether, letting $I_\text{in}$ and $I_\text{out}$ denote the number of inner and outer loop iterations and subsuming lower-order terms, the complexity of the long-term design is
\begin{align}
 \mathcal{O}\big(I_\text{out}I_\text{in} (K^2N^2+KN^3)  + I_\text{out}NK   \big).
\end{align}
In the short-term design, in turn, each iteration of the WMMSE-based approach entails $\mathcal{O}(K^2 L+KL^3)$ operations. Thus, the complexity per frame is $\mathcal{O}\big(T_\text{s} I_{\text{W}}(K^2 L+KL^3)\big)$ where $T_\text{s}$ is the number of slots and $I_{\text{W}}$ the number of WMMSE iterations per slot.

Under the premise that $N > L > K$ and $I_{\text{in}} \gg 1$, the complexity is dominated by
\begin{align} \label{CC}
\mathcal{O}\big(I_{\text{in}}I_{\text{out}}KN^3 +T_\text{s} I_{\text{W}}KL^3\big).
\end{align}
Note that the scaling with $N^3$ affects only the long-timescale optimization, which is executed once per frame rather than at every slot.


\section{Special Case: Single-User Scenario}
\label{El Pi}

The case of a single user per time-frequency resource deserves prominent attention, its formulation reducing to
\begin{subequations} \label{P_SU}
\begin{align}
& \max_{\boldsymbol{Q}}  \mathbb{E}_{\boldsymbol{g}}  \! \left[\max_{\boldsymbol{w}}~\log_2 \! \left(1+\frac{1}{\sigma^2} |\boldsymbol{g}^{*}\boldsymbol{QAw}|^2\right) \right] \label{obj_SU} \\
& \; {\rm{s.t.}}~ \eqref{cons_Q},~ \|\boldsymbol{QA}\boldsymbol{w}\|^2 \leq P_\text{t}. \label{power_constraint_SU}
\end{align}
\end{subequations}
%
Applying Algorithm \ref{Algorithm2}, $\boldsymbol{Q}$ and $\boldsymbol{w}$ are jointly optimized, but only $\boldsymbol{Q}$ is retained for use over the long term  while $\boldsymbol{w}$ is re-optimized at each slot based on the fixed $\boldsymbol{Q}$ and the effective iCSI. In contrast to multiuser settings, the single-user scenario enables a closed form for $\boldsymbol{w}$. Also, the long-term optimization of $\boldsymbol{Q}$ becomes simpler and more tractable.

\subsection{Short-Term Design}

The corresponding problem is given by
\begin{subequations} \label{P_SU_w}
\begin{align}
& \max_{\boldsymbol{w}}~|\boldsymbol{g}^{*}\boldsymbol{QAw}|^2\\
& \; {\rm{s.t.}}~ \|\boldsymbol{QA}\boldsymbol{w}\|^2 \leq P_\text{t},
\end{align}
\end{subequations}
whose solution necessarily entails equality in the power constraint, i.e., $\|\boldsymbol{QA}\boldsymbol{w}\|^2 = P_\text{t}$.
By virtue of that, and letting
$\boldsymbol{b}=\boldsymbol{A}^*\boldsymbol{Q}^*\boldsymbol{g}$ and $\boldsymbol{C}= \boldsymbol{A}^*\boldsymbol{Q}^*\boldsymbol{QA}$, the
above is equivalent to the Rayleigh-quotient maximization
\begin{subequations}
\begin{align}
& \max_{\boldsymbol{w}}~\frac{|\boldsymbol{b}^*\boldsymbol{w}|^2}{\boldsymbol{w}^*\boldsymbol{C}\boldsymbol{w}}\\
& \; {\rm{s.t.}}~ \boldsymbol{w}^*\boldsymbol{C}\boldsymbol{w}=P_\text{t},
\end{align}
\end{subequations}
whose solution is known to be
\begin{align}\label{opt_ST_w}
\boldsymbol{w}=\sqrt{\frac{P_\text{t}}{\boldsymbol{b}^*\boldsymbol{C}^{-1}\boldsymbol{b}}} \, \boldsymbol{C}^{-1}\boldsymbol{b} .
\end{align}

\subsection{Long-Term Design}

Obtaining $\boldsymbol{Q}$ entails solving
\begin{subequations} \label{P51}
\begin{align}
& \max_{\boldsymbol{Q},\boldsymbol{w}} \mathbb{E}_{\boldsymbol{g}}  \! \left[\log_2 \! \left(1+\frac{1}{\sigma^2} |\boldsymbol{g}^{*}\boldsymbol{QAw}|^2\right) \right] \\
& \; {\rm{s.t.}}~\eqref{cons_Q},~ \|\boldsymbol{QA}\boldsymbol{w}\|^2 \leq P_\text{t},
\end{align}
\end{subequations}
where the objective function is the ergodic spectral efficiency of a scalar fading channel
with average signal-to-noise ratio
\begin{align}
    \text{SNR} & = \frac{1}{\sigma^2} \mathbb{E} [|\boldsymbol{g}^{*}\boldsymbol{QAw}|^2] \\
    & = \frac{1}{\sigma^2} \boldsymbol{w}^{*}\boldsymbol{A}^{*}\boldsymbol{Q}^{*}\boldsymbol{RQAw} \\
    & = \frac{1}{\sigma^2} \| \boldsymbol{R}^{1/2} \boldsymbol{QAw} \|^2 .
\end{align}

Under the considered Rayleigh model, the single-user ergodic spectral efficiency is monotonic in the SNR. Hence, (\ref{P51}) is tantamount to
%
\begin{subequations} \label{P_SU_appr}
\begin{align}
& \max_{\boldsymbol{Q},\boldsymbol{w}} ~ \|\boldsymbol{R}^{1/2}\boldsymbol{QAw}\|^2 \\
 & \; {\rm{s.t.}}~ \eqref{cons_Q},~ \|\boldsymbol{QA}\boldsymbol{w}\|^2 \leq P_\text{t}.
\end{align}
\end{subequations}
This can be solved by the proposed PDD-based method. Defining the auxiliary variable $\boldsymbol{p}=\boldsymbol{QAw}$, the augmented Lagrangian form of \eqref{P_SU_appr} is
\begin{subequations}\label{P_AL_SU}
\begin{align}
& \min_{\boldsymbol{Q},\boldsymbol{w},\boldsymbol{p}}  -\boldsymbol{p}^*\boldsymbol{R}\boldsymbol{p}  +\frac{1}{2\rho}\left\|\boldsymbol{QAw}-\boldsymbol{p} +\rho{\bm\varphi}\right\|^2 \label{P_AL_SU_obj}\\
& \;\; {\rm{s.t.}}~ \eqref{cons_Q},~ \|\boldsymbol{p}\|^2\leq P_\text{t},
\end{align}
\end{subequations}
where $\bm\varphi$ is the dual variable and the objective function is nonconvex in $\boldsymbol{p}$. As shown in Appendix \ref{Appendix_E}, by means of the additional auxiliary variable $\boldsymbol{c}$, \eqref{P_AL_SU} can be transformed into
\begin{subequations}\label{P_AL_SU2}
\begin{align}
& \min_{\boldsymbol{Q},\boldsymbol{w},\boldsymbol{p},\boldsymbol{c}}  \boldsymbol{c}^*\boldsymbol{c}  -2\Re \big(\boldsymbol{c}^*\boldsymbol{R}^{1/2}\boldsymbol{p} \big) +\frac{ \left\|\boldsymbol{QAw} -\boldsymbol{p} +\rho{\bm\varphi}\right\|^2}{2\rho}  \label{P_AL_SU2_obj}\\
& \quad {\rm{s.t.}}~ \eqref{cons_Q},~ \|\boldsymbol{p}\|^2\leq P_\text{t},
\end{align}
\end{subequations}
which is separately convex in every optimization variable. As in the multiuser case, \eqref{P_AL_SU2} is solved in the inner loop by sequentially optimizing ${\boldsymbol{c}, \boldsymbol{p}, \boldsymbol{w}, \boldsymbol{Q}}$, with the advantage that closed forms can be found for each.

\subsubsection{Step 1}
Fixing all other variables, the problem with respect to $\boldsymbol{c}$ is unconstrained and convex, with solution
\begin{align}\label{opt_c}
\boldsymbol{c} =\boldsymbol{R}^{1/2}\boldsymbol{p}.
\end{align}
\subsubsection{Step 2}
The optimization over $\boldsymbol{p}$ can be expressed as
\begin{subequations} \label{P_v}
\begin{align}\label{v}
& \min \; \frac{1}{2\rho}\boldsymbol{p}^*\boldsymbol{p} -2\Re(\boldsymbol{p}^* \boldsymbol{d})\\
& \; {\rm{s.t.}}~ \|\boldsymbol{p}\|^2\leq P_\text{t},
\end{align}
\end{subequations}
where
\begin{align}
\boldsymbol{d}=\boldsymbol{R}^{1/2}\boldsymbol{c}+\frac{1}{2\rho}(\boldsymbol{QAw}+\rho\bm\varphi).
\end{align}
From the KKT conditions, the optimum is
\begin{align}
\boldsymbol{p} = \left( \frac{1}{2\rho}+\lambda_{\text{p}} \right)^{\!-1} \! \boldsymbol{d},
\end{align}
where $\lambda_{\text{p}} \geq 0$ is the Lagrange multiplier. If the inequality constraint is satisfied for $\lambda_{\text{p}}=0$, then
\begin{align}
    \boldsymbol{p}={2\rho}\boldsymbol{d}.
\end{align}
Otherwise, the power constraint is satisfied with equality and
\begin{align}\label{opt_v}
\boldsymbol{p} =\sqrt{P_\text{t}}\frac{\boldsymbol{d}}{\|\boldsymbol{d}\|}.
\end{align}

\subsubsection{Step 3}
The optimization over $\boldsymbol{w}$ amounts to
\begin{align}
\min_{\boldsymbol{w}}~ \left\|\boldsymbol{QAw}-\boldsymbol{p}+\rho{\bm\varphi}\right\|^2,
\end{align}
with solution
\begin{align}\label{opt_w}
\boldsymbol{w} =\left(\boldsymbol{A}^*\boldsymbol{Q}^*\boldsymbol{QA}\right)^{-1} \! \boldsymbol{A}^{*} \boldsymbol{Q}^{*} (\boldsymbol{p}-\rho\bm\varphi).
\end{align}

\subsubsection{Step 4}
The optimization over $\boldsymbol{Q}$ reduces to
\begin{align} \label{P_Q2}
& \min_{\boldsymbol{Q}}~ \left\|\boldsymbol{QAw}-\boldsymbol{p}+\rho{\bm\varphi}\right\|^2 \\
& \; {\rm{s.t.}}~ \eqref{cons_Q},
\end{align}
for which the optimum $\theta_i$ is given by \eqref{opt_q}, with
\begin{align}
\boldsymbol{B}=\boldsymbol{A}\boldsymbol{w} \boldsymbol{w}^{*} \boldsymbol{A}^{*}
\end{align}
and
\begin{align}
\boldsymbol{a}=\mathsf{diag}( \boldsymbol{A}\boldsymbol{w})^{*}(\boldsymbol{p}-\rho\bm\varphi).
\end{align}
The optimal $\boldsymbol{Q}$ follows via $\eqref{cons_Q}$.


\vspace{2mm}
In the outer loop, $\bm\varphi$ and $\rho$ are updated based on the  constraint violation, $h=\left\|\boldsymbol{QAw}-\boldsymbol{p}\right\|$. 


\section{Performance Evaluation}
\label{Sant Just}

To assess the performance, a BS equipped with a planar DMA is considered.
The carrier frequency is 28 GHz, which corresponds to $\lambda=1.07$ cm. Assuming a microstrip implemented in Duroid 5880 with 30 mil thickness \cite{Zhang2022}, the attenuation coefficient and effective dielectric constant are, respectively, $\alpha=0.6$ m$^{-1}$ and $\varepsilon=1.99$. To prevent mutual coupling, the spacing between elements is $\lambda/2$. The path loss is modeled as $10^{-3} D^{-3.5}$, where $D$ is the distance \cite[Sec. 3.3]{Lozano2018}. The users are located at $D=200$ m.

 The correlation matrices exhibit the Kronecker structure $\boldsymbol{R}= \boldsymbol{R}_{\text{H}}\otimes\boldsymbol{R}_{\text{V}}$, where $\boldsymbol{R}_{\text{H}}$ and $\boldsymbol{R}_{\text{V}}$ are spatial correlation matrices in the horizontal and vertical domains, both abiding by the exponential correlation model such that \cite[Sec. 3.6]{Lozano2018}
\begin{align}\label{correlation}
[\boldsymbol{R}_{\text{H}}]_{i,j}=r^{|i-j|}   \qquad\quad  [\boldsymbol{R}_{\text{V}}]_{i,j}=r^{|i-j|},
\end{align}
where $0<r<1$.
In the single-user case,
$r=0.5$; in the multiuser case, and with a view to representing a diverse assortment of conditions, $r_k=0.2+0.1 k$ for the $k$th user.

The noise power is set to $\sigma^2=-96$ dBm, which could correspond, e.g., to a bandwidth of 10 MHz and an 8-dB noise figure
\cite{You2023}.
(A change in $\sigma^2$ would merely shift the performance curves.)
Unless otherwise stated, $M=16$ and $L=8$ for a total of $N=128$ elements, while $K=4$. The number of slots per frame is $T_\text{s}=1000$, roughly the ratio between the rate at which channel statistics and channel realizations change in reasonable mobility conditions \cite[Sec. 3.7]{Lozano2018}.
Pilot overheads are not accounted for.

For Algorithm \ref{Algorithm2}, the initial penalty factor is $\rho=10^5$ while the dual variables are initialized to zero;
the constraint violation is $h=1$, whereas $\eta=1$ and the scaling factors are $c_1=0.5$ and $c_2=1/6$. Finally, the threshold is $\epsilon=10^{-5}$.

\subsection{Single-User Case}

The two-timescale design proposed in this paper, labeled as TTS-proposed, is evaluated against the following benchmarks:
\begin{itemize}
\item iCSI: both DMA and precoder are based on iCSI \cite{Kimaryo2023}.
\item sCSI: both DMA and precoder are based only on sCSI.
Mapped to our design, this amounts to the digital precoder obtained in Algorithm \ref{Algorithm2} being retained for the entire frame.

\item TTS-SSCA: The conventional SSCA method applied to the two-timescale problem \cite{Zhao2021, Liu2018, Cai2020, Liu2025}.

\item DMA-off: the DMA phase shifts are set to zero and the digital precoder is designed as per \eqref{opt_ST_w}.

\end{itemize}

\begin{figure}[!t]
\centering
\includegraphics[height=0.395\textwidth]{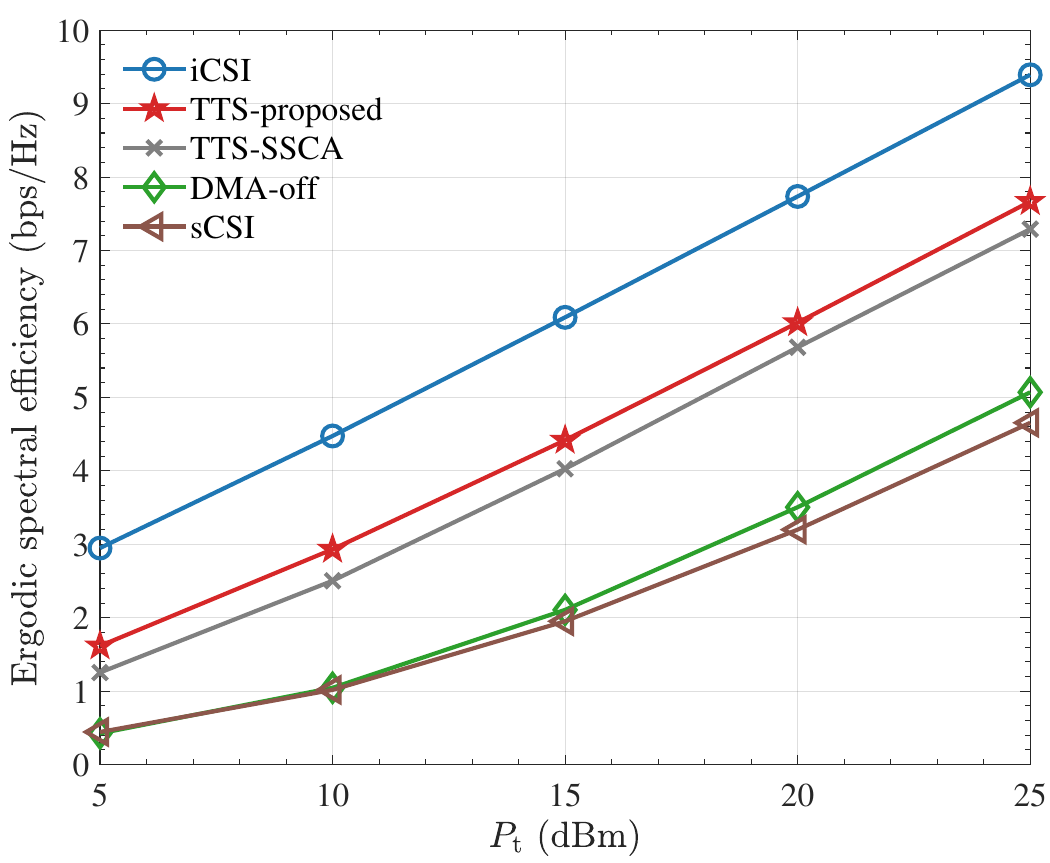}
\caption{Ergodic spectral efficiency vs transmit power.}
\label{ESR_SU}
\end{figure}

\begin{figure}[!t]
\centering

\includegraphics[height=0.393\textwidth]{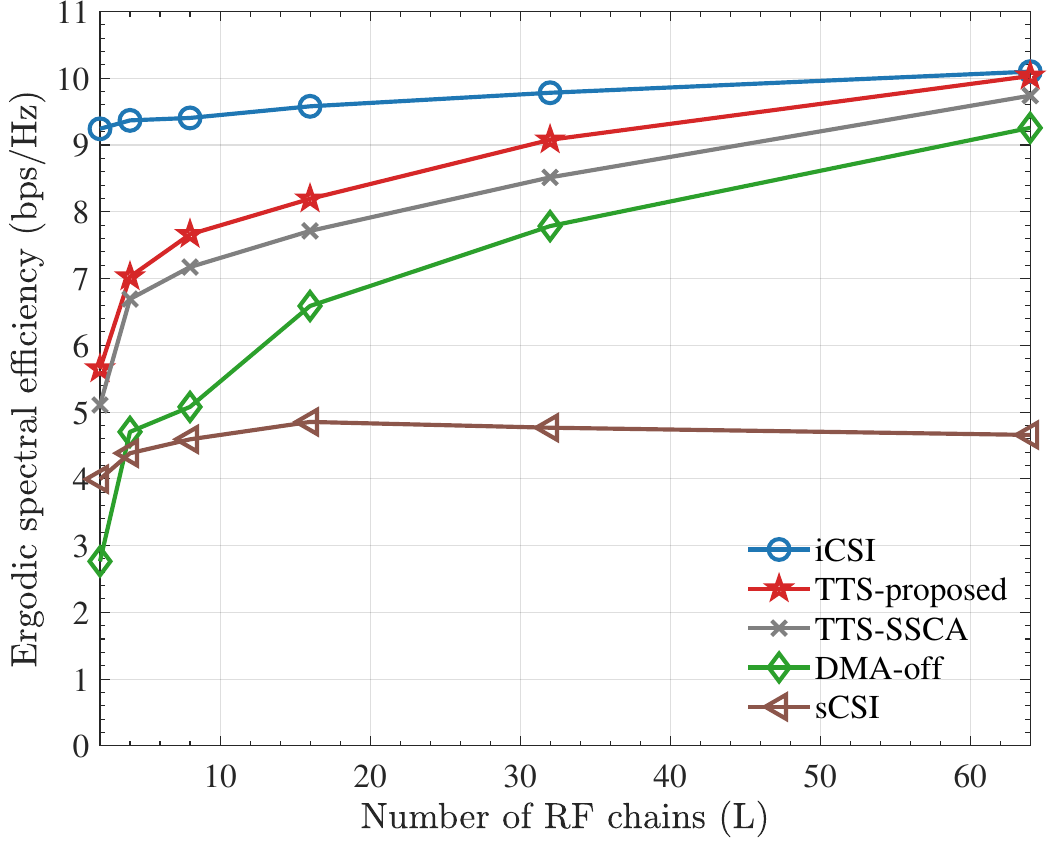}
\caption{Ergodic spectral efficiency vs number of RF chains for $P_\text{t}=25$ dBm and $N=128$.}
\label{ESR_SU_L}
\end{figure}

\subsubsection{Spectral Efficiency}

{\figurename} \ref{ESR_SU} shows the ergodic spectral efficiency as a function of the transmit power. The TTS designs go a long way towards recovering the shortfall of sCSI relative to iCSI---especially since the iCSI performance in the figure would be penalized by a much higher pilot overhead.
The TTS designs also markedly outperform the DMA-off alternative, which highlights the effectiveness of optimizing the DMA coefficients, even if only in the long term. And, between the TTS alternatives, the proposed one
exhibits a consistently superior performance.

\subsubsection{DMA Architecture}

In {\figurename}~\ref{ESR_SU_L},
the impact of the number of RF chains is examined, with the number of elements fixed. Increasing the number of chains does not enlarge the array gain, and thus this brings little improvement to the iCSI- and sCSI-based schemes. In contrast, the TTS performance improves steadily because additional RF chains enable the acquisition of further effective iCSI, improving the short-term digital precoding. As a result, the TTS designs gradually approach the iCSI performance, with the proposed one requiring substantially fewer RF chains than the SSCA-based alternative.

\subsection{Multiuser Case}

The multiuser performance evaluation relies on the same benchmarks as its single-user counterpart.

\begin{figure}[!t]
\centering
\includegraphics[height=0.393\textwidth]{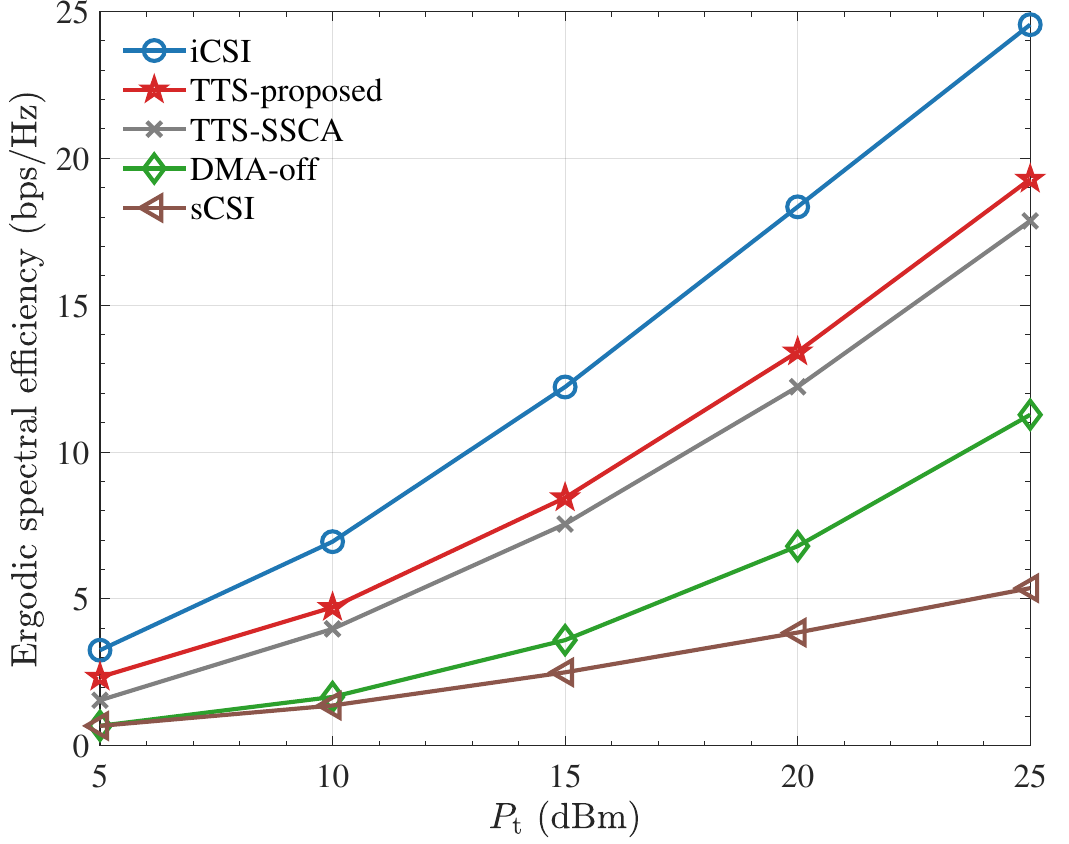}
\caption{Ergodic spectral efficiency vs transmit power.}
\label{ESR_MU}
\end{figure}

\subsubsection{Spectral Efficiency}
The observations are qualitatively similar to the single-user ones. As shown in Fig. \ref{ESR_MU}, the proposed design outperforms the TTS-SSCA alternative and it recovers a hefty share of the sCSI deficit relative to the iCSI ideal performance. Once again, we hasten to emphasize that such iCSI ideal performance would be penalized by a much higher pilot overhead, meaning that the actual recovery of the proposed TTS design is even more pronounced.

\begin{figure}[!t]
    \centering
    \subfigure
    { \hspace{0.25cm}
        \includegraphics[width=0.46\textwidth]
        {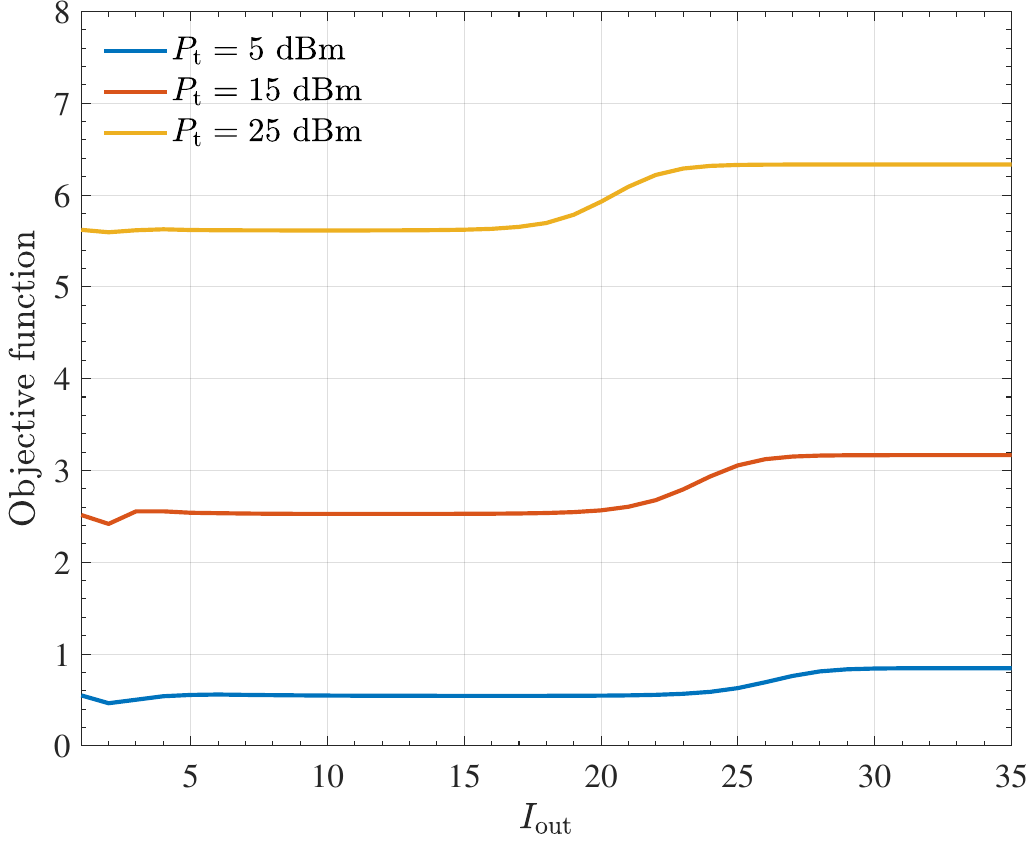}
	   \label{PDD_obj_convergence}	

    }
   \subfigure
    {
        \includegraphics[width=0.48\textwidth]{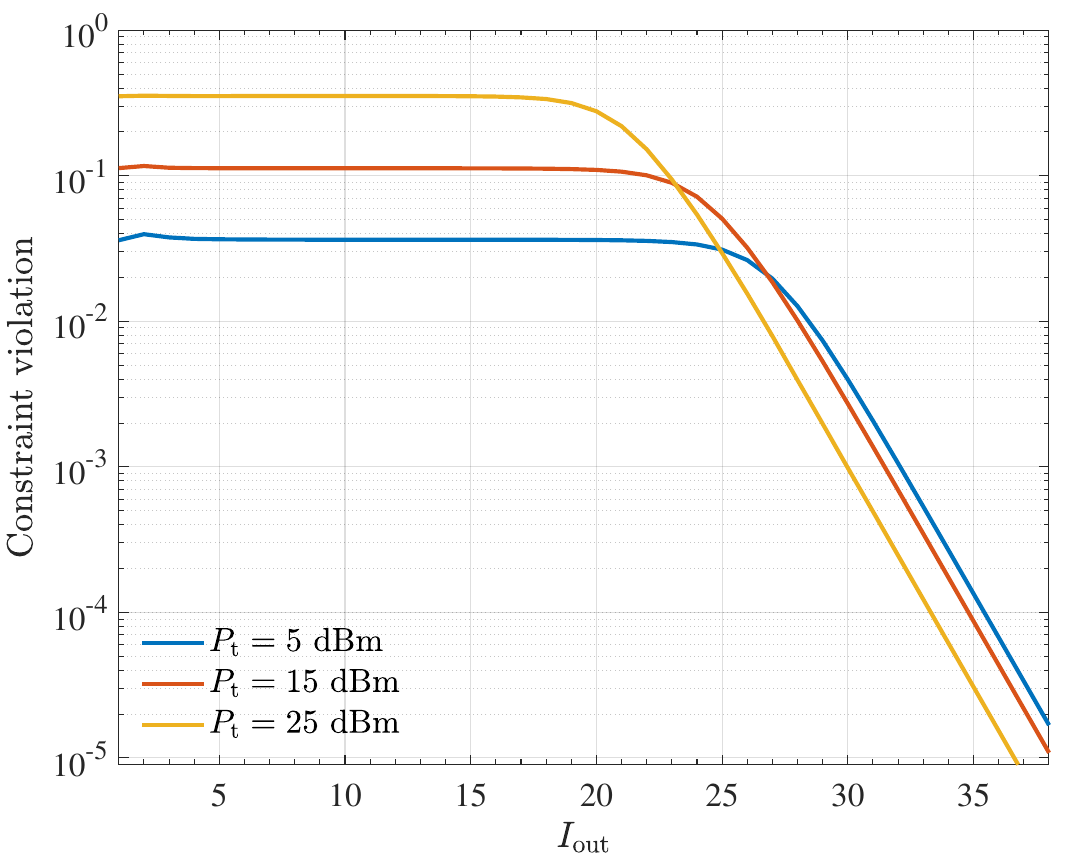}
	   \label{PDD_h_convergence}
       \hspace{0.4cm}
    }
\caption{Convergence of Algorithm \ref{Algorithm2}, parameterized by the transmit power. Above, objective function in \eqref{obj_P_appr}; below, constraint violation in \eqref{h}.
}
\label{PDD_convergence}
\end{figure}

\subsubsection{Convergence}

The convergence of the long-term optimization embodied by
Algorithm \ref{Algorithm2} is illustrated in {\figurename} {\ref{PDD_convergence}}, which shows how the objective function settles at a stationary point while the constraint violations vanish.
Regardless of the operating point, 30-40 outer iterations suffice for the performance to stabilize with only minute constraint violations.

\begin{figure}[!t]
\centering
\includegraphics[height=0.393\textwidth]{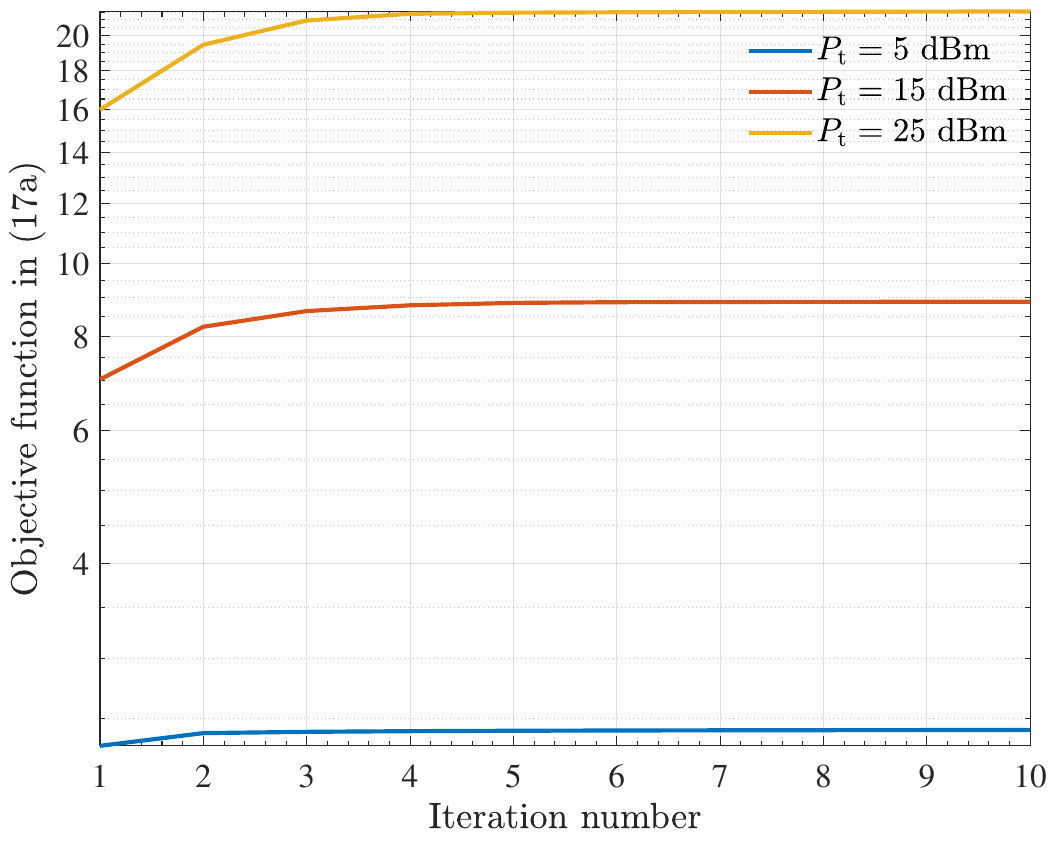}
\caption{Convergence of the short-term digital precoder.}
\label{Complexity_TTSCSI}
\end{figure}

Turning to the short-term digital precoder, the convergence of the WMMSE-based method is exemplified in {\figurename} {\ref{Complexity_TTSCSI}}.
Thanks to its decoupling from the DMA optimization and the lower dimensionality of its effective channel, the convergence is consistently fast, a couple of iterations sufficing.

\begin{figure}[!t]
\centering
\includegraphics[height=0.36\textwidth]{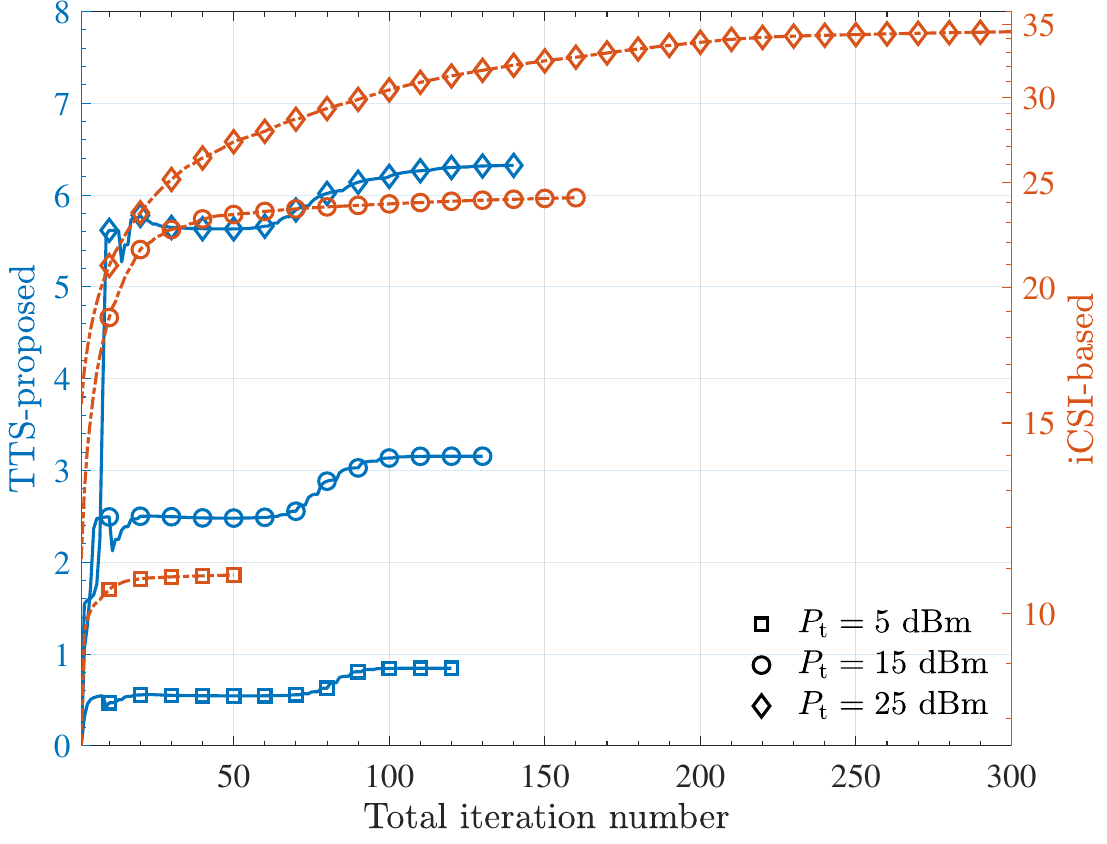}
\caption{Convergence of the iCSI-based and TTS-proposed schemes. Blue solid curves and left-side axis: objective function in \eqref{obj_P_appr} vs $I_\text{in}I_\text{out}$; red dashed curves and right-side axis: objective function of the alternating optimization vs $I_\text{o}$. 
}
\label{Complexity_iCSI}
\end{figure}


\subsubsection{Complexity}
Armed with the convergence rates of the long- and short-term optimizations, the reduction in complexity with respect to designs reliant on iCSI can be gauged.
The complexity of the proposed design, recall, is given by \eqref{CC}. In contrast, the complexity of the iCSI-based scheme is \cite{Kimaryo2023}
\begin{align}
  \mathcal{O}\Big( T_\text{s} I_\text{o} \big(I'_\text{W}KL^3 + I_\text{MO}KN^2\big)\Big) ,
\end{align}
where $I_\text{o}$ denotes the number of iterations in the alternating optimization, $I'_\text{W}$ is the number of iterations in the WMMSE-based update of $\boldsymbol{W}$, and $I_\text{MO}$ refers to the number of iterations for the manifold optimization to update $\boldsymbol{Q}$.
As evidenced by {\figurename}~\ref{Complexity_iCSI}, $I_\text{o}$ in the iCSI approach and $I_\text{in}I_\text{out}$ in the TTS-proposed approach are of the same order of magnitude. Since $I_\text{MO}\geq 1$ and $T_\text{s}I_\text{MO}\gg N$, it follows that
\begin{align}
\mathcal{O}( T_\text{s} I_\text{o} I_\text{MO}KN^2)\gg \mathcal{O}(I_{\text{in}}I_{\text{out}}KN^3).
\end{align}
In turn, the WMMSE-based optimization of $\boldsymbol{W}$ in the iCSI-based design follows a procedure akin to that in the proposed short-term design, hence $\hat{I}_\text{W} \approx I_\text{W}$, whereby
\begin{align}
\mathcal{O}\Big( T_\text{s} I_\text{o}\hat{I}_\text{W}KL^3\Big) \gg  \mathcal{O}\Big( T_\text{s}I_\text{W}KL^3\Big).
\end{align}
Altogether, the proposed design offers a decided reduction in computational burden. 

\begin{figure}[!t]
\centering

\includegraphics[height=0.393\textwidth]{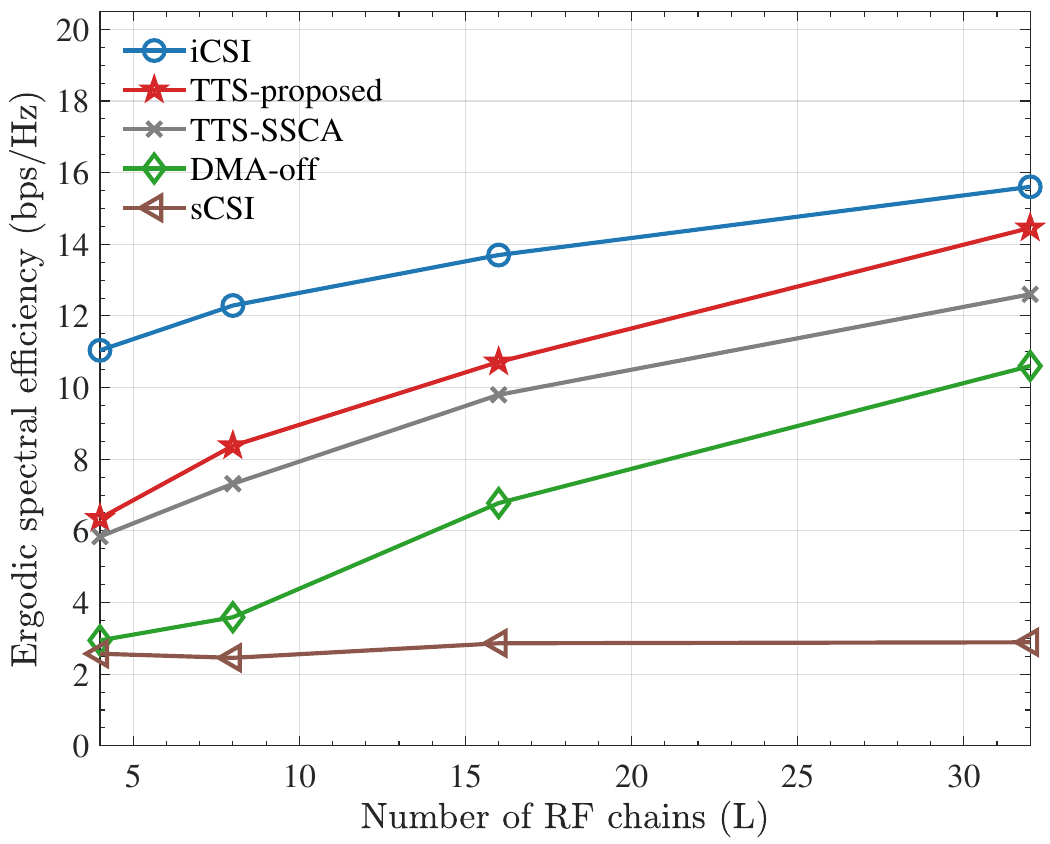}
\caption{Ergodic spectral efficiency vs number of RF chains for $P_\text{t}=15$ dBm and $N=128$.}
\label{ESR_MU_L}
\end{figure}

\subsubsection{DMA Architecture}

Illustrated in {\figurename}~\ref{ESR_MU_L} is the impact of changing the number of RF chains, with
the number of DMA elements fixed at $N = 128$. The performance gap between the iCSI-based scheme and the TTS-proposed design narrows as $L$ increases. For $L=32$, the proposed design achieves over 90\% of the ergodic spectral efficiency attained by the iCSI-based scheme, markedly outperforming the TTS-SSCA method.

\subsubsection{Validation of the Analysis}
\begin{figure}[!t]
\centering
\includegraphics[height=0.393\textwidth]{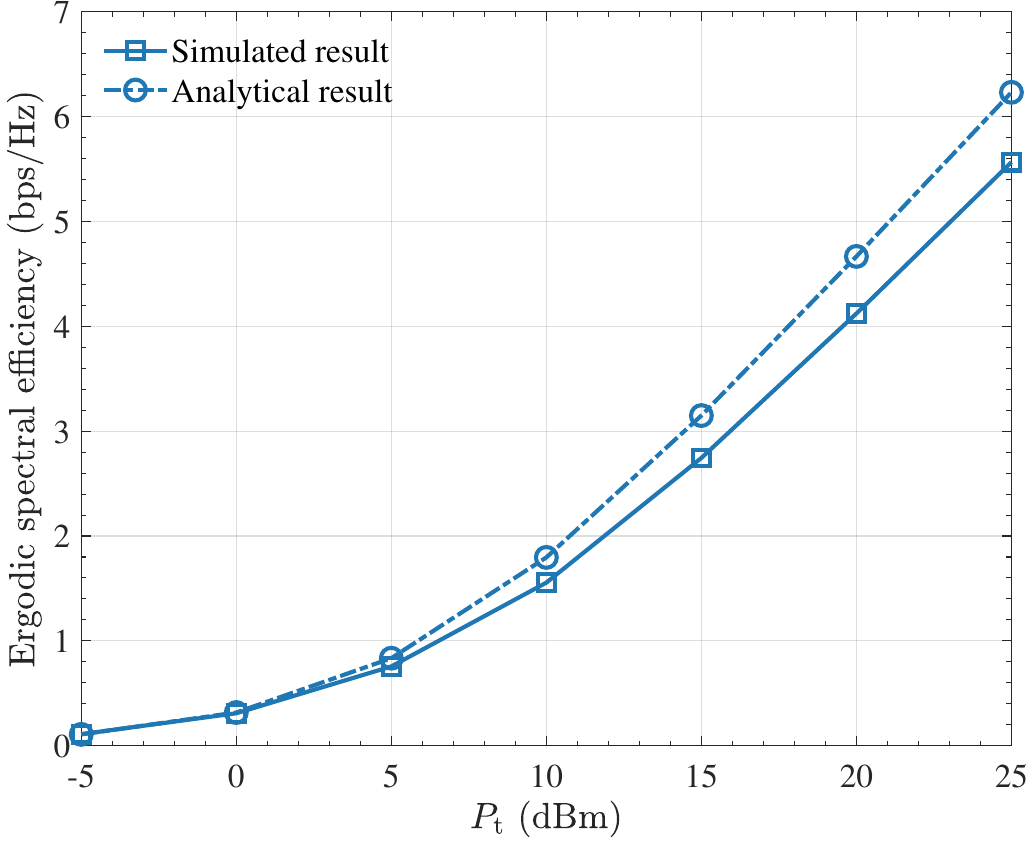}
\caption{Validation of \eqref{MU_Approximation}.}
\label{Approximation_MU}
\end{figure}

Last but not least, {\figurename} \ref{Approximation_MU} validates the approximation in \eqref{MU_Approximation}, which underpins the entire multiuser analysis. Precisely,
the figure compares the spectral efficiency obtained analytically by means of \eqref{MU_Approximation}
against the value obtained through Monte-Carlo simulation of \eqref{P_sCSI} with the same $\mathbf Q$ and $\mathbf w$. The agreement is excellent at low and medium power levels, and satisfactory at high levels, altogether validating the approach.

\section{Summary}

A two-timescale design for DMA-based downlink transmission has been presented, entailing long-term DMA coefficient adjustment and short-term digital precoding.
At the onset of each frame, the DMA coefficients are configured based only on sCSI and then, within each frame, the digital precoder is updated at each slot. 
For the DMA configuration, an optimization method that outperforms existing benchmarks has been proposed. For the digital precoding, a WMMSE-based approach has been adopted, and the precoder has been expressed in closed form for the special case of single-user transmissions.

The proposed design balances pilot overhead, computational complexity, and spectral efficiency. It bridges much of the performance gap between the sCSI- and iCSI-based schemes, yet with a computational complexity that remains  far below that of the latter.


\begin{appendix}

\subsection{Derivation of \eqref{opt_W1} } \label{Appendix_A}

Denoting the objective function in \eqref{obj_WMMSE_W} as $f(\boldsymbol{w}_k)$, the Lagrangian function associated with problem \eqref{WMMSE_W} becomes
\begin{align}\label{Lagrangian_function}
\mathcal{L}(\boldsymbol{w}_k,\lambda_{\text{w}})=f(\boldsymbol{w}_k)+ \lambda_{\text{w}} \left( \sum_{k=1}^{K}\boldsymbol{w}_k^{*}\boldsymbol{F}^{*} \boldsymbol{F} \boldsymbol{w}_k - P_\text{t}\right),
\end{align}
from whose first-order optimality condition the solution emerges as \eqref{opt_W1}.
It can be verified that, if the power constraint in the original problem \eqref{P_W} is not satisfied with equality, the SINR is lower than if the constraint is met with equality.
Thus,
\begin{align}\label{power_constraint_equal}
\text{tr}(\boldsymbol{QA}\boldsymbol{W} \boldsymbol{W}^{*} \boldsymbol{A}^{*}\boldsymbol{Q}^{*})=P_\text{t} ,
\end{align}
from which
$\lambda_{\text{w}}$ can be calculated by means of the bisection method provided that
$\text{tr}(\boldsymbol{F}\boldsymbol{W} \boldsymbol{W}^{*} \boldsymbol{F}^{*})$ is monotonic in $\lambda_{\text{w}}$.

Substituting $\boldsymbol{h}_k=\boldsymbol{F}\boldsymbol{g}_k$ into $\boldsymbol{W}$ 
yields
\begin{align}\label{W_opt2}
\boldsymbol{W}=\big( \boldsymbol{F}^{*} \hat{\bm\Omega}\boldsymbol{F} + \lambda_{\text{w}} \boldsymbol{F}^{*} \boldsymbol{F} \big)^{-1}\boldsymbol{F}^{*} \bm\Phi,
\end{align}
where
\begin{align}
\hat{\bm\Omega}=\sum_{i=1}^{K} \xi_i |u_i|^2 \boldsymbol{g}_i \boldsymbol{g}_i^{*}
\end{align}
and
\begin{align}
\bm\Phi=\boldsymbol{G} \, \text{diag}(\bm\xi)\text{diag}(\boldsymbol{u})
\end{align}
given $\bm\xi=[\xi_1,\dots,\xi_K]^{\mathsf{T}}$ and $\boldsymbol{u}=[u_1,\dots,u_K]^{\mathsf{T}}$.
Consider the thin singular value decomposition
\begin{align}
\boldsymbol{F}=\boldsymbol{UD}\boldsymbol{V}^{*} ,
\end{align}
where $\boldsymbol{U}\in\mathbb{C}^{N\times L}$ satisfying $\boldsymbol{U}^{*}\boldsymbol{U}=\boldsymbol{I}$, while $\boldsymbol{D}\in\mathbb{C}^{L\times L}$ is a diagonal matrix and $\boldsymbol{V}\in\mathbb{C}^{L\times L}$ is a unitary matrix. Substituting $\boldsymbol{F}=\boldsymbol{UD}\boldsymbol{V}^{*}$ into
$\text{tr}(\boldsymbol{F}\boldsymbol{W} \boldsymbol{W}^{*} \boldsymbol{F}^{*})$, we have that
\begin{align}
\text{tr}(\boldsymbol{F}\boldsymbol{W} \boldsymbol{W}^{*} \boldsymbol{F}^{*}) & =\text{tr}\Big((\boldsymbol{U}^{*}\hat{\bm\Omega}\boldsymbol{U}+\lambda_{\text{w}}\boldsymbol{I})^{-1} \boldsymbol{U}^{*}\bm\Phi {\bm\Phi}^{*}\boldsymbol{U} (\boldsymbol{U}^{*}\hat{\bm\Omega}\boldsymbol{U} \notag \\
& \quad +\lambda_{\text{w}}\boldsymbol{I})^{-1}\Big) .
\end{align}
Define $\boldsymbol{B}_1=\boldsymbol{U}^{*}\hat{\bm\Omega}\boldsymbol{U}\in\mathbb{C}^{L\times L}$, which is positive-semidefinite. Its eigenvalue decomposition is $\boldsymbol{B}_1=\tilde{\boldsymbol{U}}\bm\Lambda\tilde{\boldsymbol{U}}^{*}$, where $\tilde{\boldsymbol{U}}$ is a unitary matrix and $\bm\Lambda$ is a diagonal matrix with rank $R$. Plugging $\boldsymbol{B}_1=\tilde{\boldsymbol{U}}\bm\Lambda\tilde{\boldsymbol{U}}^{*}$ into $\text{tr}(\boldsymbol{F}\boldsymbol{W} \boldsymbol{W}^{*} \boldsymbol{F}^{*})$, one obtains
\begin{align}
\text{tr}(\boldsymbol{F}\boldsymbol{W} \boldsymbol{W}^{*} \boldsymbol{F}^{*})&=\text{tr}\Big((\bm\Lambda+\lambda_{\text{w}}\boldsymbol{I}_L)^{-2} \boldsymbol{C}_1\Big)\\
&=\sum_{i=1}^{R}\frac{[\boldsymbol{C}_1]_{i,i}}{( [\bm\Lambda]_{i,i} +\lambda_{\text{w}})^2},
\label{marroc}
\end{align}
where $\boldsymbol{C}_1=\tilde{\boldsymbol{U}}^{*} \boldsymbol{U}^{*}\bm\Phi {\bm\Phi}^{*}\boldsymbol{U} \tilde{\boldsymbol{U}}$ and \eqref{marroc} is seen to be monotonically decreasing in $\lambda_{\text{w}}$.
Thus, the bisection method can be applied.

\subsection{Proof of \eqref{MU_Approximation} } \label{Appendix_B}
Letting
\begin{align}
A_k
&=
\sum_{i=1}^{K}
\boldsymbol w_i^* \boldsymbol A^* \boldsymbol Q^*
\boldsymbol g_k \boldsymbol g_k^*
\boldsymbol Q \boldsymbol A \boldsymbol w_i
+\sigma_k^2,
\\
B_k
&=
\sum_{i\neq k}
\boldsymbol w_i^* \boldsymbol A^* \boldsymbol Q^*
\boldsymbol g_k \boldsymbol g_k^*
\boldsymbol Q \boldsymbol A \boldsymbol w_i
+\sigma_k^2,
\end{align}
the objective function can be written as
\begin{equation}
\sum_{k=1}^{K}\mathbb{E} \big[\log_2 A_k -\log_2 B_k \big].
\end{equation}
By virtue of Jensen's inequality,
\begin{align}
\mathbb{E} \big[\log_2 A_k \big]&\le\log_2 \mathbb{E}[A_k],
\\
\mathbb{E} \big[ \log_2 B_k \big]&\le\log_2 \mathbb{E}[B_k],
\end{align}
whereby, in line with similar Jensen-based approximations \cite{Li2012,Gan2021}, the objective function can be approximated by
\begin{equation}  \label{derive_R}
\sum_{k=1}^{K} \Big(
\log_2 \mathbb{E}[A_k] -\log_2 \mathbb{E}[B_k] \Big).
\end{equation}
Using $\mathbb{E}[\boldsymbol g_k\boldsymbol g_k^*]=\boldsymbol R_k$, furthermore
\begin{align}
\mathbb{E}[A_k]
&=
\sum_{i=1}^{K}
\boldsymbol w_i^* \boldsymbol A^* \boldsymbol Q^*
\boldsymbol R_k
\boldsymbol Q \boldsymbol A \boldsymbol w_i
+\sigma_k^2,
\\
\mathbb{E}[B_k]
&=
\sum_{i\neq k}
\boldsymbol w_i^* \boldsymbol A^* \boldsymbol Q^*
\boldsymbol R_k
\boldsymbol Q \boldsymbol A \boldsymbol w_i
+\sigma_k^2.
\end{align}

\subsection{Derivation of \eqref{P_FP} } \label{Appendix_C}

The Lagrangian dual transform is applied here \cite{Shen2018_2}. By introducing the auxiliary variables $\{\Gamma_k\}$, \eqref{P_appr} can be transformed into
\begin{subequations} \label{P_Ld}
\begin{align}
\!\!  \max_{\boldsymbol{Q},\boldsymbol{W},\{\Gamma_k\}} & \sum_{k=1}^{K} \log(1+\Gamma_k)-\sum_{k=1}^{K}\Gamma_k \label{obj_Ld}\\
&+\sum_{k=1}^{K}(1+\Gamma_k) \frac{\boldsymbol{w}_k^{*} \boldsymbol{A}^{*}\boldsymbol{Q}^{*} \boldsymbol{R}_k \boldsymbol{QA}\boldsymbol{w}_k }{\sum_{i=1}^{K}\boldsymbol{w}_i^{*} \boldsymbol{A}^{*}\boldsymbol{Q}^{*} \boldsymbol{R}_k \boldsymbol{QA}\boldsymbol{w}_i+\sigma_k^2} \nonumber \\
{\rm{s.t.}}&~ \eqref{cons_Q}, ~\sum_{k=1}^{K}\|\boldsymbol{QA}\boldsymbol{w}_k\|^2\leq P_\text{t}.
\end{align}
\end{subequations}
From the first-order optimal condition it follows that $\Gamma_k=\bar{\gamma}_k$,
which, plugged back into \eqref{obj_Ld}, yields the objective function in \eqref{obj_P_appr}. Problems \eqref{P_appr} and \eqref{P_Ld} thus have the same optimal objective value and the same optimal solution for $\boldsymbol{Q}$ and $\boldsymbol{W}$. This establishes the equivalence between the two problems.

To decouple the numerator and denominator of the fraction in \eqref{obj_Ld}, one can further transform \eqref{P_Ld} into \eqref{P_FP} by introducing the auxiliary variables $\{\bm\beta_k\}$ and resorting to the quadratic transform technique \cite{Shen2018_1}. The equivalence argument is similar to the one above, hence it is omitted for brevity.

\subsection{Derivation of \eqref{P_AL_SU2}} \label{Appendix_E}

Denote the objective functions in \eqref{P_AL_SU_obj} and \eqref{P_AL_SU2_obj} by $f_1(\boldsymbol{Q},\boldsymbol{w},\boldsymbol{p})$ and $f_2(\boldsymbol{Q},\boldsymbol{w},\boldsymbol{p},\boldsymbol{c})$, respectively. From the first-order optimal condition, the optimal solution for $\boldsymbol{c}$ in \eqref{P_AL_SU2_obj} is $\boldsymbol{c}=\boldsymbol{R}^{1/2}\boldsymbol{p}$. Inserted into $f_2(\boldsymbol{Q},\boldsymbol{w},\boldsymbol{p},\boldsymbol{c})$, this yields
\begin{align}
f_2(\boldsymbol{Q},\boldsymbol{w},\boldsymbol{p}, \boldsymbol{c})=f_1(\boldsymbol{Q},\boldsymbol{w}, \boldsymbol{p}) ,
\end{align}
indicating that problems \eqref{P_AL_SU} and \eqref{P_AL_SU2} have the same optimal objective value and the same optimal solution for $\{\boldsymbol{Q},\boldsymbol{w}, \boldsymbol{p}\}$.
\end{appendix}

\end{document}